\documentclass[conference]{IEEEtran}

\usepackage[numbers,sort&compress]{natbib}

\usepackage{caption}
\usepackage{subcaption}
\usepackage{times}
\usepackage{array}
\usepackage{footnote}
\usepackage[hyphens]{url}
\usepackage{xspace}
\usepackage{setspace}
\usepackage{tablefootnote}
\usepackage[binary-units]{siunitx}
\usepackage{textcomp}   % for special math symbols (such as \mu) used by siunitx
\usepackage[dvipsnames]{xcolor}

\usepackage{cryptocode}

\newcolumntype{R}{>{\centering}X}

\usepackage{amsmath,amssymb,amsfonts}
\usepackage{hyperref}
\usepackage{cleveref}
\usepackage{graphicx}
\usepackage{mathtools}
\usepackage{multirow}
\usepackage{clrscode3e}
\usepackage{algorithm}
\usepackage{algorithmic}
\usepackage{booktabs} % to thicken table lines
\usepackage{enumitem} % to remove indent for second line of \item ([wide] option)

\newlength\myindent
\newcommand{\Hide}[1]{}

\Crefname{alg}{Algorithm}{Algorithms}	% let Cref know name for algorithm environment

\newcommand{\BfPara}[1]{\smallskip \noindent{\bf #1}.}

\newcommand{\initiator}[0]{\mathcal{I}}     % initiator
\newcommand{\reflector}[0]{\mathcal{R}}     % reflector
\newcommand{\adversary}[0]{\mathcal{A}}
\newcommand{\epoch}[0]{\tau}                   % time epoch
\newcommand{\key}[0]{K}                     % shared key
\newcommand{\batch}[0]{\mathcal{B}}
\newcommand{\bestss}[0]{\mathcal{S}}
\newcommand{\var}[1]{\mathit{#1}}           % a variable with 2 or more characters
\newcommand{\fn}[1]{\texttt{\small #1}}           % a function name
\newcommand{\sync}[0]{\var{SYNC}}
\newcommand{\req}[0]{\var{REQ}}
\newcommand{\resp}[0]{\var{RESP}}
\newcommand{\argmax}[0]{\mathrm{argmax}}

\makeatletter

\def\old@comma{,}
\catcode`\,=13
\def,{%
  \ifmmode%
    \old@comma\discretionary{}{}{}%
  \else%
    \old@comma%
  \fi%
}
\makeatother

\hypersetup{draft}

\title{Spectrum-Flexible Secure Broadcast Ranging}

\begin{document}

%%% Authors IEEE template
\author{
    \IEEEauthorblockN{Tien D. Vo-Huu}
    \IEEEauthorblockA{\textit{Northeastern University} \\
    Boston, USA \\
    tienvh@ccs.neu.edu}
    \and
    \IEEEauthorblockN{Triet D. Vo-Huu}
    \IEEEauthorblockA{\textit{Northeastern University} \\
    Boston, USA \\
    vohuudtr@ccs.neu.edu}
    \and
    \IEEEauthorblockN{Guevara Noubir}
    \IEEEauthorblockA{\textit{Northeastern University} \\
    Boston, USA \\
    g.noubir@northeastern.edu}
}

\maketitle

\graphicspath{{figures/}}

\begin{abstract}
Secure ranging is poised to play a critical role in several emerging applications such as self-driving cars, unmanned aerial systems, wireless IoT devices, and augmented reality. In this paper, we propose a design of a secure broadcast ranging systems with unique features and techniques. Its \textit{spectral-flexibility}, and \textit{low-power short} ranging bursts enable co-existence with existing systems such as in the 2.4GHz ISM band. We exploit a set of RF techniques such as \textit{upsampling} and \textit{successive interference cancellation} to achieve high accuracy and scalability to tens of reflectors even when operating over narrow bands of spectrum. We demonstrate that it can be implemented on popular SDR platforms FPGA and/or hosts (with minimal FPGA modifications). The protocol design, and cryptographically generated/detected signals, and randomized timing of transmissions, provide stealth and security against denial of service, sniffing, and distance manipulation attacks. Through extensive experimental evaluations (and simulations for scalability to over 100 reflectors) we demonstrate an accuracy below 20cm on a wide range of SNR (as low as 0dB), spectrum 25MHz-100MHz, with bursts as short as 5us. 
\end{abstract}

\section{Introduction}

The mobile revolution, fundamentally changed how we access and share information. Wireless localization has greatly contributed to this revolution, by enabling access to geographically relevant information (e.g., Location Based Services). It is poised to have an even greater and more critical impact in the future of navigation systems, from self-driving cars, to air traffic coordination, to unmanned aerial systems (e.g., delivery drones), but also for a growing number of IoT and Augmented Reality applications. 

Secure localization received an increased interest from the research community in recent years, as practical attacks were demonstrated against a variety of system~\cite{rasmussen07secnav,TippenhauerPRC2011,RanganathanC2017,OlafsdottirRC2017}. For instance, spoofing GPS signals is easily achievable with open source software~\cite{gps-sdr-sim} and lowcost SDR platforms. This led to demonstration of hijacking drones~\cite{NohKSSKCK2019}, but also the occurrence of real world incidents were detected in multiple places~\cite{gps-spoofing-new-scientist,gps-spoofing-maritime}. Augmenting GPS with Inertial Navigation System has proven to be insecure too~\cite{NarainRN2019}. Localization systems for air traffic also  proved to be insecure with vulnerabilities including landing an airplanes off the runway by attacking the ILS~\cite{SathayeSRN2019}, or inserting a ghost airplane in the sky exploiting weaknesses in  ADS–B~\cite{CostinA2012,StrohmeierSPLM2017}. 

Ranging is an important building block of localization systems as it enables two devices to estimate their relative distance. Various techniques were proposed and deployed over the years. These include energy-based ranging, phase-based ranging, and time-based ranging. The accuracy of most these techniques and systems is expected to be at least 1 meter~\cite{zafari2017survey}. However, existing techniques, including the recent IEEE 802.15.4 addition of Scrambled Timestamp Sequence (STS), remain vulnerable to various attacks. In essence, cryptographic techniques alone are insufficient to defend against spoofing and relaying signals without a common trusted time reference, and interactive protocols are subject to jamming and sophisticated replaying of signals by adversaries, if such signals can be detected. In the security research community one class of protocols focused on distance-bounding protocols where a party proves that it is within a given distance from a verifier. Distance-bounding protocols are important for various access control applications. Proposed techniques aim at mitigating distance-decreasing attacks such as  early-detect/late-commit (ED/LC)~\cite{rasmussen08db,poturalski11db,manuel10dd}, and denial of service (DoS)~\cite{poturalski10cicada}. Other protocols were proposed to protect the time of arrival (ToA) and enhance the integrity of the ranging information~\cite{rasmussen07secnav,tippenhauer09temp}, and more recently a security primitive for securing messages ToA derivation~\cite{leu20mtac}.

In this work, we propose a set of integrated techniques and protocols to secure broadcast ranging. The proposed techniques are designed to operate over flexible spectrum bandwidth and can be implemented both on an FPGA and/or host device without performance loss. For this we exploit both sub-sampling  and scheduled burst-mode capabilities of modern RF SDR peripherals. The techniques and protocols provide both anti-jamming protection and support of multi-device (broadcast) ranging thanks to stealthy cryptographically randomized messages in time and code, as well as successive interference cancellation. We extensively evaluate the performance of the proposed techniques, analytically, with simulations for scalability (to tens of simultaneous reflectors), and experimentally over-the-air on SDR platforms. We in particular explore the trade-offs in terms of spectrum usage, interference resilience, accuracy, and scalability. Our techniques are implemented on the USRP X310~\cite{wiki:usrpX310}, a popular Software Defined Radio (SDR) platform, both on the FPGA and host using GNU Radio~\cite{url:gnu-radio} and our own extension of  UHD~\cite{wiki:uhd} for a fine control of the added capabilities to the X310 FPGA. We demonstrate flexibility to operate over a spectrum of 25-100MHz, achieving below 20 cm accuracy with bursts of 5 us. Our contribution are summarized as follows:

\begin{itemize}
	\item We develop, to our best knowledge, the first SDR ranging system with secure broadcast mechanism at high accuracy on a flexible narrow bandwidth.
	\item Stealth and security are achieved through cryptographically randomized messages in time and code.
	\item Flexibility and scalability are achieved through upsampling and successive interference cancellation.
	\item The protocol is designed and implemented on FPGA, host (with minimal FPGA modifications), and hybrid.
	\item We demonstrate through extensive performance evaluations (over-the-air and simulations for scalability) that we can achieve an accuracy below 20cm on a wide range of SNR (as low as 0dB), spectrum 25MHz-100MHz,  20 simultaneous reflectors (with sessions of $\SI{100}{us}$), and over 100 reflectors with sessions of $\SI{1}{ms}$.
\end{itemize}
\section{Background and Related Work}
\label{sec:background}

\subsection{Ranging Techniques Overview}
Most popular ranging techniques fall into one of three main types: energy-based, phase-based, and time-based estimation.

\BfPara{Energy-based Ranging} 
The core idea of energy-based ranging relies on the assumption that the relationship between the distance and the Received Signal Strength can be expressed by a simple path loss model. While this model provides an easy and inexpensive method to compute the distance, it usually achieves poor accuracy (in order of meters) even in free-space environments due to the complex wireless channel propagation and fading effects and irregular antenna pattern. As a result, it is typically only used in low-cost systems, where complex signal processing cannot be afforded~\cite{zanella16rssranging}.

\BfPara{Phase-based Ranging} When a signal is transmitted on a certain frequency, its phase change is a periodic function of the travel time.
Transmitting signals on a set of different frequencies and measuring the received phases, the distance can be derived unambiguously. This is the basis for phase-based ranging typically used in multitone communication systems.
Since clock synchronization is not required between nodes, phase-based ranging is suitable for many applications. For instance in Wi-Fi systems, previous work has demonstrated that decimeter accuracy could be achieved~\cite{vasisht16chronos,manikanta15spotfi}.

\BfPara{Time-based Ranging} In contrast to phase-based ranging, the time-based ranging technique relies on the transmission and reception time of a signal, and the distance is computed based on the Time of Flight (ToF)~\cite{dardari09}. Time-based ranging systems require high precision of ToF measurements. For example, a small error of \SI{10}{\ns} can result in a large distance error of \SI{3}{\m}.
Without time synchronization, the ranging between nodes is typically performed in a two-way fashion, where a return signal is sent back to help estimate the ToF independently of the time reference at the other end.
With the recent introduction of ranging capability into the IEEE 802.15.4 standard~\cite{ieee802154}, ranging systems based on Impulse Radio Ultra Wide Band (IR-UWB) become popular, such as Decawave~\cite{decawave}, 3dB Access~\cite{3dbaccess}.
Ranging using SDR has been recently explored in \cite{sark17usrpranging}. In their work, by using USRP N210 with maximum rate of \SI{50}{MHz} together with the use of a maximum-length sequence (m-sequence) with length of $1024$ for detection, they could achieve \SI{40}{\cm} accuracy.

\subsection{Ranging Attacks}
\label{sec:background-attacks}

\BfPara{Sniffing} While messages exchanged between ranging nodes can be encrypted to hide the embedded private data, a sniffer can infer the distance based on its observation of the transmitted signal in the wireless medium. The observation can be focused on the change in amplitude or phase, and derive an estimate of ToF. While deriving sub-metre accuracy is challenging, achieving metre level accuracy might satisfy the adversary scenarios. Furthermore, the attack success rate can be improved if a capable adversary is in proximity.

\BfPara{Denial of Service}
With DoS attacks, the goal of the attacker is to prevent or degrade the ranging session. To perform such attacks, jamming signals are emitted and targeted to the communication link between benign nodes. Depending on the underlying physical layer used for ranging, jamming does not have to be continuous. An example of non-continuous jamming attacks is the cicada attack~\cite{poturalski10cicada}, which targeted to the IR-UWB ranging by the use of intermittent pulses that can block or degrade the distance estimation at the receiver.

\BfPara{Distance Enlargement}
In this attack, an adversary aims to deceive the ranging system into thinking that the nodes are farther than the actual distance~\cite{Francillon2010RelayAO}. As a result, the system could deactivate protection mechanisms e.g., collision avoidance in self-driving cars or drones. The feasibility of these attacks depends on the adversary's capability of guessing the victim's signals in advance and adaptively generating malicious signals. The adversary can also replay the transmitted signal with overshadowing or annihilation as demonstrated in IR-UWB systems~\cite{taponecco13,singh19uwbed}.

\BfPara{Distance Reduction}
As the opposite of distance enlargement, the goal of distance reduction attacks is to decrease the perceived distance, often between entities who are out of their communication range, thus enabling the adversary to bypass the distance-based security system (e.g., PKES, smart home security). The basis of this attack relies on the signal relaying capability of the man-in-the-middle adversary, who is able to send a possibly modified version of the request or response such that a shorter distance is resolved. This attack is specially effective against energy-based or phase-based ranging systems, since the replay version can be crafted with an appropriate amplifier factor or phase change~\cite{olafsdottir2017security}.

For time-based ranging in a theoretic setting, it is impossible to perform distance reduction attacks due to the nature of relaying that increases the ToF. Based on this principle, the distance bounding protocol~\cite{brandchaum01db} and its variants~\cite{brelurut} were developed as a logical layer on top of an existing ranging physical layer to provide the protection from distance reduction attacks. In practice, however, if the distance bounding protocol is not integrated properly, the adversary can still perform these attacks by exploiting the properties of the concrete physical layer employed in the system without breaking any cryptographic assumptions made by the distance bounding protocol. One physical-layer attack against distance bounding protocol is Early-Detect/Late-Commit (ED/LC), which exploits the latency in the modulation and demodulation in RFID radios~\cite{hancke08rfid}, or pulses characteristics of both preamble and payload in IR-UWB~\cite{manuel10dd}, or in chirp-based system~\cite{aanjhan12cbattack}. 
The main idea of the attack is to take advantage of the predictable signal to transmit a guessed portion of the signal earlier than the ranging node and fill the rest immediately after some legitimate part is revealed. In UWB, the early-detection phase corresponds to the guess of the whole symbol when only half of symbol is revealed, and the late commit phase corresponds to the filling of higher pulses or zeros to compensate for any incorrect guess. To prevent this attack, a recent countermeasure has been proposed for UWB that employs the pulse reordering method~\cite{singh19uwbpr}. 
\section{Proposed Protocol: Overview}
\label{sec:problem-approach}
Our approach is time-based, because it is easier to secure (requiring more efforts to overcome to timing constraints) than energy-based and phase-based ranging.

\subsection{Threat Model}
\label{sec:threat-model}
The system consists of multiple nodes executing the ranging protocol at any given time to learn their distance. The main purpose of an external adversary is \emph{either to learn the private distance information, to block the ranging, or to deceive the nodes into perceiving a wrong distance}. 

\BfPara{Cryptographic Capabilities}
Ranging nodes use a shared key $\key$ for cryptographic operations. We assume the key $\key$ is shared using an out-of-band channel. Messages exchanged for ranging are produced using either a symmetric encryption or a cryptographic pseudorandom function (PRF) with the shared key $K$. The cryptographic operations we use are assumed to be secure and cannot be broken by the adversary.

\BfPara{Communication Capabilities}
The communication parameters, such as center frequency, bandwidth, modulation, coding are public, and the adversary is able to capture and transmit signals on the same channel as the ranging nodes. The adversary can create interference with the ranging transmissions by transmitting its own signal. However, as messages are derived using secure cryptographic operations, the adversary cannot predict the transmitted signal (indistinguishable from random), and consequently he is unable to annihilate or modify the signal in a meaningful way to himself. We also assume that the adversary cannot transmit signal with higher or lower propagation speed compared to other transmitters in the same wireless medium. This assumption implies that a signal replay will increase the ToF.

\BfPara{Mobility}
Nodes are assumed to be mobile, and at the time of a ranging session, a node has no knowledge whether others are in its communication range. Similarly the adversary does not know the exact location of the ranging nodes. However, the adversary may know whether ranging nodes are in their communication range based on side channel information.

\BfPara{Attack Scenarios} We consider all the attack scenarios discussed in~\Cref{sec:background}, namely sniffing, denial of service, distance enlargement, and distance reduction attacks. In all scenarios, we assume honest (non-malicious) ranging nodes in the presence of an adversary.

\subsection{Approach Overview}
\label{sec:approach}

\begin{figure*}[thb]
    \centering
    \includegraphics[width=0.8\textwidth]{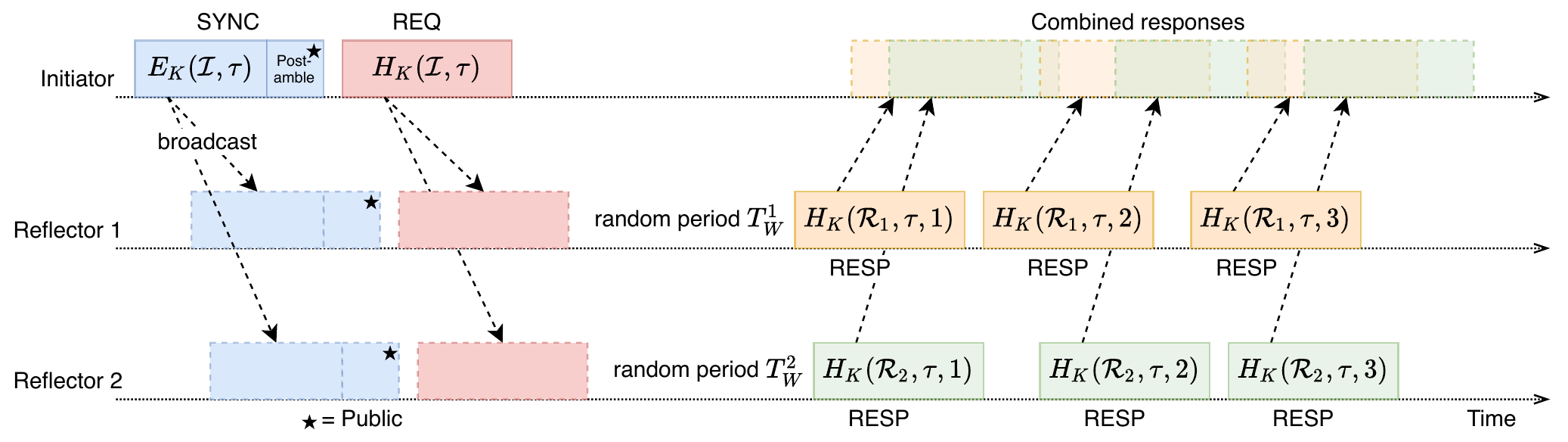}
    \caption{Example of Secure broadcast ranging protocol between one initiator and two reflectors. The SYNC message is sent only if nodes are unsynchronized. The postamble is marked with a black star as the only public content in the protocol. The initiator broadcasts the REQ and expects to receive responses as a combined signal (collision) from reflectors. Both REQ and RESP messages are secret symbol sequences without a preamble (in contrast with the SYNC). In this example, each reflector replies with 3 responses. The waiting period between the REQ and the first RESP and between consecutive RESPs  are cryptographically randomized.}
    \label{fig:multidelay}
\end{figure*}

\begin{table}[t]
    \small
    \centering
    \begin{tabular}{cl}
    \toprule
        $\initiator$        & Initiator node starting a ranging session \\
        $\reflector_k$      & Reflector node sending back response to initiator\\
        $\key$            & Shared key used by ranging nodes \\
        $E_{\key}(m)$   & Encryption of message $m$ using key $\key$ \\
        $H_{\key}(m)$    & Secure PRF of $m$ using key $\key$ (e.g., HMAC-SHA2) \\
        $\epoch$            & Current time epoch \\
        $\Delta{}\epoch$    & Time epoch duration \\
        $t_S$          & Send time of the request recorded by initiator \\
        $t_R$          & Received time of the response recorded by initiator \\
        $T$                 & Sampling period \\
        $T_E$               & Subsample timing error \\
        $T_W$               & Random waiting period between request and response \\
        $T_{RESP}$          & Fixed duration of a ranging response \\
        $W$                 & Waiting window\\
        $\batch$            & A batch of responses replied per reflector\\
    \bottomrule
    \end{tabular}
    \caption{Notations.}
    \label{tab:notations}
\end{table}
We consider a ranging session between multiple nodes, in which one node, called the \emph{initiator} $\initiator$, is interested in discovering the distance from itself to other nodes, called \emph{reflectors}\footnote{From the security perspective, the initiator is known as the \emph{verifier}, who initiates the secure ranging protocol and makes judgement about the distance computed based on responses from the reflectors that play the role of the private information \emph{prover}. However, this terminology is used when the prover is not necessarily honest.} $\reflector_1,\reflector_2,\ldots$ Our notations are summarized in~\Cref{tab:notations}. The ranging session is started by the initiator.
In our ranging protocol, we introduce the \emph{time epoch} as the discrete time instant computed from the real-time system clock as $\epoch=\lfloor{}t/\Delta{}\epoch\rfloor$ with $t$ and $\Delta{}\epoch$ representing the current real time and the period between two consecutive epochs, respectively. 
Our ranging protocol, illustrated in~\Cref{fig:multidelay}, is a time-based ranging protocol consisting of the following steps.

\begin{enumerate}[wide] % <-- this option removes indent from second line of item
    \item \emph{Synchronization:}
The main goal of this step is to ensure the nodes in our system are loosely synchronized with respect to the time epoch. It is \emph{not} required for every ranging session. To perform the synchronization, the initiator transmits a $\sync$ message consisting of the initiator's identifier and current time epoch $\epoch$ IND\$-CPA  encrypted (indistinguishable from random) with the shared key $\key$~\cite{Rogaway2004}.
\begin{equation}
\label{eq:sync}
\sync := \{E_{\key}(\initiator,\epoch) | \var{Postamble}\}.
\end{equation}
The synchronization message is embedded in a frame with a public postamble that enables the reflectors to receive it.

    \item \emph{Requesting:}
The ranging procedure is started by the initiator choosing a random epoch and sending out a ranging request $\req$ at the beginning of the epoch. If a prior $\sync$ message is required, $\req$ is sent right after that. The $\req$ message is a fixed-length sequence of physical signal symbols generated with a cryptographic PRF, shared key $\key$, the initiator's identifier, and current time epoch $\epoch$. 
\begin{equation}
\label{eq:req}
\req := H_{\key}(\initiator,\epoch).
\end{equation}
In contrast to $\sync$, no public preamble or postamble is used for $\req$. On the other side, as long as $\epoch$ is synchronized (based on a previous synchronization), the reflectors begin scanning for the secret sequence.

    \item \emph{Responding:}
Once receiving the ranging request, each reflector $\reflector_k$ will send back a response batch $\batch$ consisting of multiple $\resp^{k}_{n}$ messages ($n=0\ldots{}|\batch|-1$), which are secret symbol sequences hashed with the shared key $\key$.
\begin{equation}
\label{eq:resp}
\resp^k_{n} := H_\key(\reflector_k,\epoch,n). \\
\end{equation}
Each response is transmitted at an exact time specified by a random waiting period $T_W$, a secret value generated per the reflector using a cryptographic PRF and shared key (on reflector identifier, $\tau$, and response counter). The use of multiple responses and random waiting periods by each reflector not only improves the accuracy of distance computation for the ranging session, but also increases the challenges to the adversary. The ranging session is ended when the reflectors complete all $|\batch|$ responses, and finally the initiator computes  the distance to each of them.
\end{enumerate}

We now highlight key properties of our protocols. A more thorough discussion is presented in~\Cref{sec:protocol,sec:security}.

\BfPara{Secure Ranging}
A major reason for successful exploits (e.g., ED/LC attacks) on the physical layer of existing ranging systems is rooted from the adversary's knowledge of preambles prepended to the request and response exchanged between nodes. Using public preamble is, in fact, the typical design of today's existing communication systems for the purpose of signal synchronization and channel estimation. In our ranging protocol, the request and response are indistinguishable from noise, and  are detectable and decodable only by ranging users. This effectively provides protection against attacks based on early preamble detection. We also note that the synchronization is not a part of the distance computation process, and it is infrequently transmitted.

The use of secret sequences of random symbols not only shields us from being detected, but also mitigates the impact of jamming attacks. This essentially enables our ranging system to operate under low signal-to-noise (SNR) conditions.
The use of random waiting period $T_W$ prevents the adversary from learning the actual distance, even with coarse accuracy.

One can notice that our protocol is asymmetric (i.e., no feedback from the initiator after the responses). As a result, the reflectors are unable to estimate the distance to the initiator with a good accuracy. This is our intended design from the security perspective, which protects the private distance information from honest-but-curious reflectors. For applications where this security feature is not required, our ranging protocol can be extended (e.g., by making the protocol symmetric or using the higher layer for exchanging information afterwards) to disclose the distance to reflectors without the need of initiating a new ranging session from each reflector.

\BfPara{Broadcast Ranging}
Our ranging protocol does not put constraints on the order of responses, which can come back at the initiator in form of a collision. The collision, however, does not prevent the initiator from learning the distance, but enables saving communication bandwidth for the whole system while increasing challenges to the adversary. At the initiator, individual responses from each reflector can be extracted from the combined received signal thanks to the independence of encrypted symbol sequences. In addition, we use a Successive Interference Cancellation (SIC) technique to improve the receptability and accuracy of responses from far-away reflectors under the interference from close-by ones. 

\BfPara{Accuracy}
With the two-way ranging procedure, the initiator can unlock itself from the dependency on the reflectors' clock. However, the accuracy of distance estimation tightly depends on the sample period (higher sampling rate results in better accuracy). In this work, we build our ranging prototype on USRP platform and operate the system under \SI{100}{\MHz}. This gives the timing error of $\SI{\pm 5}{\ns}$ (equivalent to distance error of $\SI{\pm 1.5}{m}$), making sample rate the bottleneck in achieving high accuracy of ToF estimation. To overcome this issue, we use a subsampling interpolation technique to achieve subsample ToF estimation and significantly improve the accuracy.

\BfPara{Spectrum Flexibility}
The need for large bandwidth for accurate ranging (e.g., IW-UWB with \SI{500}{\MHz}) limits the application of existing ranging protocols, as many communication systems operate in limited spectrum (e.g., ISM band). In contrast, our ranging system can operate in an \emph{upsampling} mode, where the signal bandwidth is much smaller than the sample rate, limiting interference to other communications.
\section{Secure Broadcast Ranging Protocol}
\label{sec:protocol}

In this section, we provide a detailed description of our ranging protocol with the focus on ranging accuracy. As a time-based ranging protocol, the distance computation in our protocol relies on the precision of recording the requesting and responding time. While sharing the basic mechanism with existing time-based ranging systems (such as IR-UWB), where the Time of Flight is measured to compute the distance, our underlying physical layer processing techniques are different and  achieve comparable accuracy with a \emph{significantly smaller bandwidth} (\SI{100}{MHz} and below).
We dedicate \Cref{sec:security} for the security analysis of the protocol. 

We first consider a simple ranging session consisting of a single request and a single response. For simplicity of presentation, we assume both initiator and reflector know the sequences used for request and response. We present the timing challenges and techniques to estimate the distance with sub-meter accuracy. We then describe the synchronization procedure used to establish the requirements for the upcoming ranging session. Finally, a full ranging session consisting of multiple responses that offer both higher accuracy and better security is presented. This section is finally ended with discussion bandwidth efficiency improvements.

\subsection{Narrowband Sub-meter Ranging}
\label{sec:distance-computation}

\begin{figure*}
    \centering
    \includegraphics[width=0.8\textwidth]{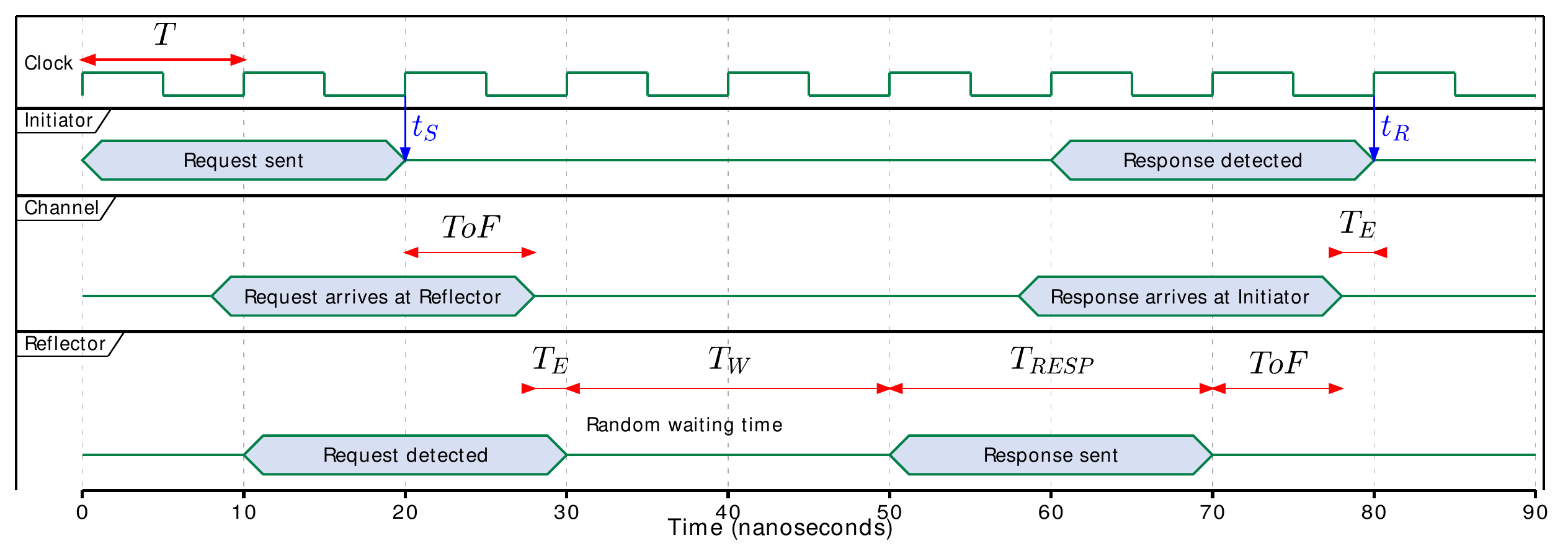}
    \caption{Timing analysis of an example ranging session carried out in bandwidth of $\SI{100}{MHz}$ ($T=\SI{10}{ns}$). Signals are sampled at positive clock edges. The initiator records the sent time $t_S$ at the request's last symbol and the received time $t_R$ at the response's last symbol. The secret random waiting period $T_W$ and response duration $T_{RESP}$ are known to the initiator. The ToF is computed by the initiator based on the $\var{RTT}=t_R-t_S$ and the estimation of timing error $t_E$.}
    \label{fig:timing}
\end{figure*}

\subsubsection{Timing Analysis}
\label{sec:timing-analysis}
Consider the ranging session consisting of one request-response pair, where an initiator $\initiator$ starts a ranging request and a reflector $\reflector$ replies with a response (\Cref{fig:timing}). The start time $t_S$ is recorded as the sent time of the last symbol of the request, and the received time $t_R$ is recorded at the last symbol of the response. Both $t_S$ and $t_R$ are based on the initiator's clock. As such, the distance is computed solely based on the time recordings done at the initiator. The reflector's time is not used for the distance computation, whereas the random waiting period $T_W$ and the response duration $T_{\var{RESP}}$ are known to the initiator.

% \BfPara{Clock resolution}
For the ranging purpose, we are interested in the precise timing of the transmission. Since digital samples are transmitted on clock edges, the recorded value $t_S$ is the actual sent time. The recorded received time $t_R$, however, might not reflect the actual arrival time of the message, as signals are only sampled at discrete time instants with a sample period $T$ while the signals can arrive anytime within the period. The difference between the recorded and actual arrival time is represented by the timing error $T_E$. A positive value of $T_E$ indicates that the actual arrival time is earlier than the recorded receive time, whereas $T_E$ is negative in case the signal is early sampled. Base on the timing diagram in~\Cref{fig:timing}, the Round Trip Time (RTT) is formulated as:
\begin{equation}
\label{eq:timing1}
\var{RTT} = t_R-t_S = 2(\var{ToF} + T_E) + (T_W + T_{\var{RESP}})
\end{equation}
In \Cref{eq:timing1}, we have made an assumption that the ToF is considered symmetric between nodes (the travel time is the same from either side) during the very short ranging duration. The ToF can be derived as
\begin{equation}
\label{eq:timing2}
ToF=\frac{1}{2}(t_R-t_S-T_W-T_{\var{RESP}}) - T_E
\end{equation}
It is worth noting that while a clock drift might exist between the initiator and reflector (typically up to $\pm\SI{40}{ppm}$), it is negligible in comparison with the timing error $T_E$, which can be as large as the sample period ($-T<T_E<T$). Therefore, $T_W$ and $T_{\var{RESP}}$ are considered deterministic and identically observed on both sides. The unknown and dominant factor determining the accuracy of ToF is the timing error $T_E$, which will be addressed in the following discussion.

\subsubsection{Time of Flight Estimation}
\label{sec:tof-estimation}
The Time of Flight estimation involves the initiator detecting the response to obtain the recorded received time $t_R$ and timing error $T_E$. 
Our signal detection method relies on the computation of cross-correlation between received signal and known response signal. The ranging request and responses are secret random sequences of complex symbols generated as in~\Cref{eq:req,eq:resp}. They possess good correlation properties (\Cref{sec:sequences}) such that the correlator outputs with high probability a large correlation value, when the pattern is present in the received signal, while producing significantly smaller value for noise or an uncorrelated sequence. Not only the presence of the pattern can be detected, the position of the peak also determines the received time instant $t_R$. 
Since the cross-correlation is energy-sensitive, we use its normalized version (relatively to the received signal energy) to enable detection in extreme cases of low and high Signal to Interference plus Noise Ratio (SINR). The normalized cross-correlation $C_l$ computed on a $L$-length sequence at a time lag $l$ is given by 
\begin{equation}
\label{eq:xcorr-norm}
	C_l =
	\frac{\sum_{n=0}^{L-1} r_{n-l} p_n^*}
	     {\sqrt{(\sum_{n=0}^{L-1}|r_{n-l}|^2) (\sum_{n=0}^{L-1}|p_n|^2)}}
\end{equation}
where $*$ denotes the complex conjugate operator, and $\{r\}$, $\{p\}$ are the received samples and the pattern to be detected, respectively. The pattern is located at position $M$, where the peak of the absolute value of $C_l$ is found.
\begin{equation}
\label{eq:peak}
    t_R = M = \argmax_l{}|C_l| \quad \textrm{if } C_l\ge\alpha C_{l'}, \quad l\ne{}l', |l-l'|\le L_0
\end{equation}
In~\Cref{eq:peak}, a peak is the local maximum of cross correlation that is significantly higher than the values in its vicinity $[l-L_0,l+L_0]$ by a ratio threshold $\alpha$. Setting $\alpha$ and $L_0$ appropriately reduces the false positive detection rate.
\BfPara{Timing Correction}
\label{subsample}
The peak computed based on~\Cref{eq:xcorr-norm,eq:peak}, however, only approximates the true peak, which could be found if we were able to sample the signal exactly at  arrival time. In reality, the received time $t_R$ might be recorded earlier or later than the actual arrival time with  average  error $E[T_E]=T/2$.
For example, if our system is sampling at \SI{100}{\MHz} (i.e., $T=\SI{10}{\ns}$), without additional technique the average distance error would be $\SI{1.5}{\m}$. 

To improve the accuracy, we estimate the timing error $T_E$ by interpolating the cross-correlation to achieve the subsample peak. The subsampling interpolation is based on the model that the vicinity of the true peak on the cross correlation curve can be approximated by an analytical function. Essentially, the true peak can be computed based on a few adjacent points in its surrounding region. This technique is typically used in digital ultrasonic measurement systems, where the distance to a target body (e.g., sea ground) is measured by emitting a signal and estimating the RTT from the passively reflected signal. The average timing error in this case is half of that in comparison to in our digital ranging system, because there is no timing error at the passive reflector in ultrasonic measurement systems. In this work, we considered several approximation functions for interpolation~\cite{svilainis13subsample,svilainis2008analysis,viola2005spline} and found Gaussian function achieves the best balance between accuracy improvement and computational complexity.

Specifically, if $C_M$ denotes the cross-correlation at the peak found in~\Cref{eq:peak}, the discrete-time values around $C_M$ can be modelled by a Gaussian function $C(x)=a\cdot{}\exp\left(-\frac{(x-b)^2}{2c^2}\right)$ for $x\in[M-1,M+1]$. Using the points $C_{M-1},C_M,C_{M+1}$, we can compute the true peak and derive the the timing error as
\begin{equation}
    T_{E} = -T \frac{\ln{C_{M+1}} - \ln{C_{M-1}}} {4 \ln{C_{M}} - 2 \ln{C_{M-1}} - 2 \ln{C_{M+1}}}.
\end{equation}

\subsection{Synchronization}
\label{sec:sync}
In the distance computation presented in~\Cref{sec:distance-computation}, the discussion assumes a that the request and response sequences are known in advance to both participant nodes. In our protocol, this condition is realized by synchronization. 

\BfPara{Loose Synchronization}
The initiator performs the synchronization step at the beginning of its current time epoch $\epoch$. When receiving the $\sync$ message, given by \Cref{eq:sync}, the reflector updates its current epoch to $\tau$. If the processing delays at the initiator and reflector are $\epsilon_{\initiator}$ and $\epsilon_{\reflector}$, the mismatch between the initiator and reflector's clocks after the epoch update is $\Delta=\epsilon_{\initiator}+\epsilon_{\reflector}+ToF+T_E$. Due to $\Delta$, the initiator and reflector are not perfectly aligned. Nevertheless, if the epoch period is large enough ($\Delta\epoch\gg\Delta$), synchronization allows both sides to be loosely synchronized with respect to the time epoch and agree on the request and response sequences used for the upcoming ranging session. In fact, the mismatch $\Delta$ is in order of a few microseconds, while the epoch period $\Delta\epoch=\SI{1}{second}$ in our system. We emphasize that  ranging accuracy is not impacted by $\Delta$, as the reflector's clock is not used for the distance computation. 

\BfPara{Infrequent Synchronization}
As long as nodes are loosely synchronized, this state remains valid in many subsequent time epochs. Synchronization messages are not necessary for each ranging session. While there are various methods to optimize how often the synchronization should be performed, our simplified approach is to schedule periodic synchronizations such that the clocks are not mismatched by more than half of the time epoch duration. Specifically, if $\delta$ is the clock drift (time difference per second) between nodes, resynchronization is required after $T_{\var{resync}}=\frac{1}{2}\Delta\epoch/\delta$ seconds. With a typical value of $\delta=\pm\SI{40}{ppm}$, we only need to synchronize roughly every $3.5$ hours. 
It is possible that a node can sometimes miss the synchronization message (e.g., due to environment conditions). When this situation is detected, for instance, being unsynchronized for significantly more than half an epoch, resynchronization can be triggered. 

\BfPara{Postamble}
The $\sync$ message has a special structure, in which the end of the message is appended by a postamble. The postamble is a sequence with a good correlation property allowing easy signal detection. In contrast of a preamble, the postamble prevents the payload of the $\sync$ message from being attacked, because when the adversary detects the $\sync$, the data is already received. The tradeoff is the increased processing overhead at the receiver, to store the whole message before decoding. As $\sync$ is a short frame, we consider this tradeoff worth the improved resiliency.

\subsection{Detailed Protocol Description}
\label{sec:protocol-details}

In this section, we present a detailed protocol description in a generic setting, where one initiator starts ranging with \emph{multiple} reflectors and learns the distance to each of them \emph{at the same time}. Broadcast ranging is a distinct feature of our protocol in comparison with existing work. To enable broadcast ranging, we allow the reflectors to send back responses that can overlap both in time and spectrum. In our protocol, each ranging session is required to start and end in the same time epoch, and each epoch serves at most one session. Within an epoch, the request is sent at the beginning, while responses are sent back at random moments. For compact presentation, we assume nodes are already loosely synchronized, i.e., they observe the same time epoch $\epoch$.

\subsubsection{Random Sequences}
\label{sec:sequences}
Random sequences are a crucial part of our ranging protocol, as they allow ranging nodes to detect requests and responses, as well as to derive timing estimation.
We focus on the correlation property of secret random sequences, as this property determines the detection capability of the system. As generated in~\Cref{eq:req,eq:resp}, a secret random sequence is obtained using a PRF (e.g., HMAC-SHA2/3), shared key, the current time epoch $\epoch$ and the sender's identifier. For the responses, we also embed an additional field for the counter $n$ to distinguish the messages. The generated sequences are modulated using BPSK modulation to create the transmitted signals. With this construction, every signal is unique and detectable to the ranging nodes, but remains indistinguishable to the adversary.

\begin{figure}[h]
	\includegraphics[width=0.47\textwidth]{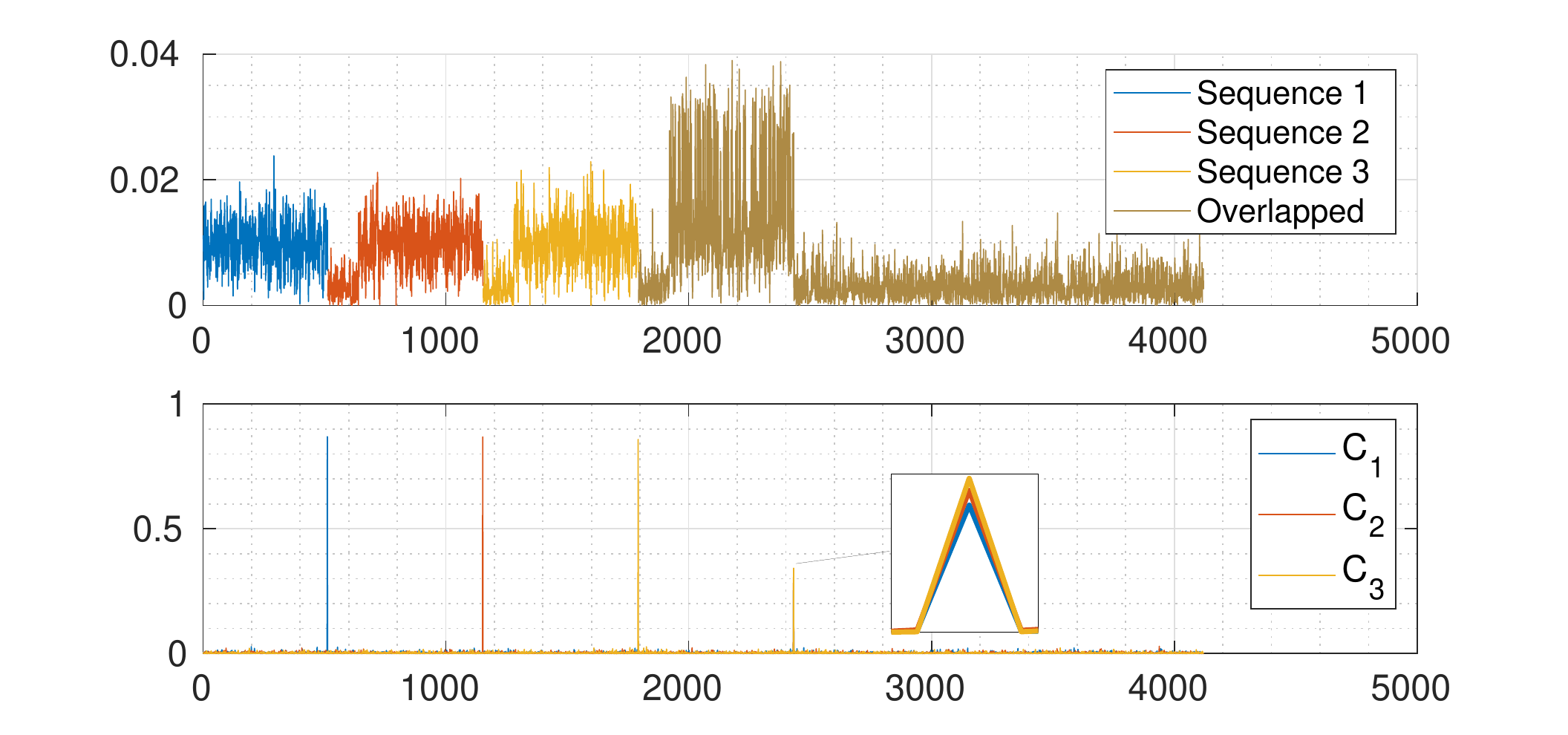}
	\caption{Example of 3 different sequences transmitted sequentially (non-overlapped), and simultaneously (overlapped). The non-overlapped region shows sharp and high peaks at the detected  position of each sequence. The overlapped region shows lower peaks, but they are still strong and clearly detectable from interference.}
	\label{fig:overlap-sequences}
\end{figure}

\BfPara{Good Correlation Property}
In terms of communication robustness, a sequence with good correlation property can be easily detected even under the presence of interference and noise. While there are good sequences with this property (e.g., m-sequences), they are not indistinguishable to the adversary from the security perspective. The security concern is our main motivation for using cryptographic operations to generate the sequences. Due to the randomness, both ranging request and response signals span the whole operating spectrum. With multiple reflectors responding at the same time, these responses potentially create signal collision at the initiator. The detection of individual responses from the combined received signal relies on the correlation property of the responses. We carry out an evaluation, where random sequences are transmitted in two scenarios: non-overlapping and overlapping. An example in~\Cref{fig:overlap-sequences} shows that our cryptographically randomized sequences possess a good correlation property that enables the initiator to easily locate individual responses under external interference and noise.

\subsubsection{Boosting Interference Resilience}
As seen from~\Cref{fig:overlap-sequences}, while our random sequences are well uncorrelated, the detection accuracy becomes lower when ranging is performed with more reflectors, because individual SINRs are reduced due to mutual interference. The impact is more visible to far-away reflectors whose signal power is already weak arriving at the initiator.
To mitigate the impact of mutual interference, we use a Successive Interference Cancellation technique to increase the SINR of individual responses. Our technique consists of the following steps.
\begin{enumerate}[wide]
\item \emph{Finding strongest reflector:} We compute the peaked correlation $C^k_M$ for each reflector $\reflector_k$ and find the strongest reflector $\reflector_{\hat{k}}$ with the highest correlation value: $\hat{k}=\argmax_k |C_M^k|$. We estimate the submeter-accuracy distance for $\reflector_{\hat{k}}$ by performing the timing correction described in~\Cref{sec:tof-estimation}.
\item \emph{Estimating channel attenuation:}
The channel attenuation $\gamma$ (complex value) for reflector $\reflector_{\hat{k}}$ is estimated as

$$\gamma=\frac{\sum_{n=0}^{N-1} r_{n+M} p_n^*}{\sum_{n=0}^{N-1}|p_n|^2}$$ 
where $\{p_n\}$ is the strongest reflector $\reflector_{\hat{k}}$'s response sequence, and $\{r_{n+M}\}$ is the received signal aligned to $\{p_n\}$ at the peak position $M$  found in the previous step.
\item \emph{Removing strongest response:} With the channel attenuation $\gamma$, the strongest response signal arriving at the initiator is estimated to be $\{\gamma p_n\}$. We can remove it from the combined received signal and repeat iteratively on the residual signal to extract remaining reflectors. Our evaluation (\Cref{sec:evaluation}) shows accuracy improvements of broadcast ranging.
\end{enumerate}

\subsubsection{Response Batch Mode}
In the following paragraphs, we discuss the response batch mode that supports multiple responses per reflector in the responding step. The main idea is to use random delays between responses to obfuscate the actual ToF. The response batch not only increases the ranging resiliency/security, but improves the ranging accuracy as well.

\BfPara{Response Delay}
We consider the responding step performed by a reflector $\reflector_k$ when it receives the initiator's request. Based on the cross-correlation computed during the request detection, the reflector starts its own timer at the peak position where the request is detected. The $n$-th response is transmitted when the timer reaches the waiting period $T^k_{W,n}$, given by
\begin{equation}
\label{eq:waiting}
T^k_{W,n} := T\cdot(H_\key( \reflector_k,\epoch,n,W) \bmod{W})
\end{equation}
where $W$ denotes the waiting window.
The timer is reset when the sending of a response is complete, then the next response is scheduled.
By~\Cref{eq:waiting}, the exact transmit time is determined uniquely for each reflector and each response, but it is non-deterministic to the adversary.
Recall the ToF estimation in~\Cref{eq:timing2}, the random waiting period results in obfuscated ToF. This prevents the adversary from learning the actual distance even with a coarse accuracy.

\BfPara{Improved Accuracy with Batch Estimation}
With the response batch $\batch$ from a reflector $\reflector_k$, the initiator can perform the distance computation and obtain multiple estimations. It is noted that the actual response batch detected by the initiator may have less responses due to missed or corrupted messages. If $\tilde{\batch}$ denotes the received batch, we have $|\tilde{\batch}|\le|\batch|$. To derive the final distance result to $\reflector_k$, the initiator first finds the subset $\bestss$ of those responses whose timing estimations are closest to the median estimation of $\tilde{\batch}$, then computes the final result as the expected value of the subset $\bestss$. The size of $\bestss$ is selected as $|\bestss| =\min(|\batch|/2, |\tilde{\batch}|$.
This estimation method filters out the outliers.

\subsubsection{Spectrum Flexibility}
Digital ranging systems in general need a large bandwidth for achieving a good accuracy. According to the Cramer-Rao bound, the accuracy of timing estimation is a function of the spectrum bandwidth used for ranging. One would achieve better accuracy with larger bandwidth. This bound implies limited accuracy for narrowband wireless ranging. In practice, however, the accuracy is far from the theoretical bound due to imperfections such as channel variations, obstructions, measurement errors~\cite{dardari09,zanella16rssranging}. 

\begin{figure}
    \centering
    \includegraphics[width=0.47\textwidth]{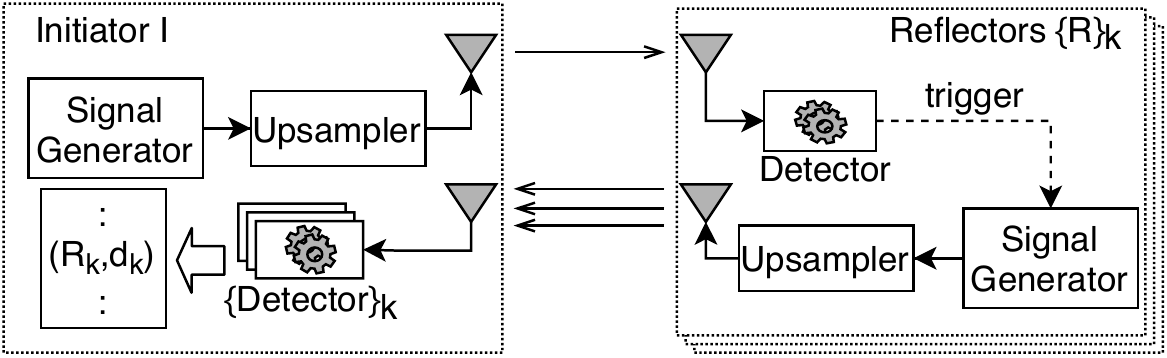}
    \caption{Ranging with upsampling mode for spectrum flexibility.}
    \label{fig:system-diagram}
\end{figure}

For our proposed ranging protocol, we find that  accuracy is rather constrained by clock resolution, which in turn depends on the sample rate. The implies  that as long as a specific sample rate yields an acceptable accuracy for a ranging application, we can shrink the bandwidth while keeping the sample rate unchanged without decreasing the accuracy. This is the motivation for the \emph{upsampling mode} in our ranging protocol (\Cref{fig:system-diagram}). To flexibly narrow down the bandwidth, both request and response signals are upsampled by an interpolation factor equal to $f_S / B$, where $f_S$ is the sample rate, and $B$ is the desired signal bandwidth. The upsampled signals are sent to the RF front end for transmission. 

In contrast to the transmitting chain, the receiving chain of both initiator and reflectors does not include a downsampler. Instead, the expected sequence is also upsampled and directly used by the detector that performs the pattern searching at the high sample rate $f_S$. This design allows us to achieve good accuracy with much narrower bandwidth and distinguishes our ranging system from previous work.

\section{Security Analysis}
\label{sec:security}
In this section, we analyze the security of our proposed protocol with the focus on the attacks introduced in~\Cref{sec:background-attacks}, and threat model described in~\Cref{sec:threat-model}.

%%%%%%%%%%%%%%%%%%%%%%%%%%%%%
\subsection{Denial of Service Attack}
Denial of Service (DoS) in ranging systems is generally carried out in the form of a jamming attack with the adversary emitting interfering signals to corrupt the ranging messages. In practice, a jammer with powerful capabilities (such as high power, continuous jamming) would be easily detected. Therefore, we consider a bounded adversary with limited jamming power similar to ranging nodes. 

We first consider a blindly intermittent jamming (e.g., cicada attack). By the use of random sequences, the SINR of the request and response signals is increased by roughly $L$ times with $L$ being the sequence length. For instance with $L=1024$, the SINR is improved by approximately $\SI{30}{dB}$. The high resilience against interference is verified in our evaluation (\Cref{sec:evaluation}), which implies their robustness against jamming attack. For $\sync$ messages, due to their infrequent use, the hit probability remains low in comparison with other transmissions in the system. For systems having ranging as an additional feature, this attack would be more efficient if it focused on non-ranging messages.

Consider a selective jamming attack, where the adversary targets the ranging messages. To perform this attack, message detection is required. However, $\req$ and $\resp$ are indistinguishable by the adversary, and the $\sync$ can only be detected at the end of the message (due to the postamble) but too late for jamming.

%%%%%%%%%%%%%%%%%%%%%%%%%%%%%
\subsection{Distance Enlargement Attack}
We consider the distance enlargement attack carried out by a man-in-the-middle adversary. In this attack, the adversary aims to delay the $\req$ and $\resp$ messages (the $\sync$ message is not used for distance computation). Since the adversary cannot generate a legitimate random sequence, delaying these messages needs to be done by recording a transmission and replaying it at a later time. However, simply replaying a message would be detected by a duplicate check at the ranging nodes. A successful distance enlargement attack, therefore, would require both blocking and replaying the ranging messages at the same time. In the following, we investigate the possibility of this attack.

First, we emphasize that blocking and recording a signal are mutually exclusive operations. In addition, as the adversary cannot predict the random sequence, blocking a signal is performed in the form of overshadowing (jamming) rather than annihilation. We consider a scenario, where the adversary records a portion of the signal from the beginning, then blocks the rest. Concretely, let $\{p_1\ldots{}p_L\}$ be a transmitted response sequence. Assume the adversary records the portion $\mathcal{X}=\{p_1\ldots{}p_N\}, N\le{}L$, and blocks the rest by a jamming signal $\{q_1\ldots{}q_{L-N}\}$. Note that $\mathcal{X}$ can be used as (a part of) the jamming signal. The (partially) blocked signal arriving at the initiator is $\mathcal{Y}=\{p_1\ldots{}p_N,p_{N+1}+q_1\ldots{}p_L+q_{L-N}\}$. We state that replaying $\mathcal{X}$ succeeds if $\mathcal{X}$ is detected while $\mathcal{Y}$ is not detected by the initiator. 

We note that the earliest time for replaying $\mathcal{X}$ is right after the recording, i.e., the first $N$ symbols of the jamming signal can be the recorded portion: $\{q_1\ldots{}q_N\}=\mathcal{X}$. Now if $ToF_{\max}$ is the maximum ToF allowed in the ranging system, $\mathcal{X}$ cannot be replayed later than $N_{\max}=ToF_{\max}/T$ samples, or the initiator will detect the attack. This requirement limits the size of $\mathcal{X}$ to $N_{\max}$, that is $|\mathcal{X}|\le{}N_{\max}$. Recall the cross-correlation computation in~\Cref{eq:xcorr-norm}, the peak correlation value of $\mathcal{X}$ would be $|\mathcal{X}|/L\le{}N_{\max}/L$ times lower than the peak of a legitimate sequence. In a real-world ranging system, if the allowed distance limit is $\SI{300}{m}$, the sample rate is $T=\SI{100}{MHz}$, and sequence length $L=1024$, we have $N_{\max}=100$ and $N_{\max}/L<1/10$. This ratio is too low for $\mathcal{X}$ to be detected.

Consider the condition for $\mathcal{Y}$ to be not detected. Recall the good correlation property of our random sequences, the jamming portion is uncorrelated with the overshadowed portion of $\mathcal{Y}$. Therefore, as long as the jamming power is of same order as ranging signals, the initiator would detect $\mathcal{Y}$ with high probability (spreading gives $\approx30dB$ gain). In conclusion the probability for a successful replay for the distance enlargement purpose is significantly low.

%%%%%%%%%%%%%%%%%%%%%%%%%%%%%
\subsection{Distance Reduction Attack}
To successfully carry out a distance reduction attack, the adversary must be able to guess the signal and relay it earlier than the legitimate one. However, since ranging responses in our protocol are indistinguishable, guessing the signal with probability higher than a random guess is impossible. Therefore, similar to the distance enlargement, the distance reduction requires the adversary to record the signal. This requirement  implies the signal (or any part of it) cannot be replayed earlier than its recording time, or equivalently our protocol is robust against this attack.

%%%%%%%%%%%%%%%%%%%%%%%%%%%%%
\subsection{Sniffing Attack}
A passive attacker aims at learning the distance to and between ranging nodes needs to detect the signal and infer the distance without the knowledge of secrets used by ranging nodes. We first see that with the use of encryption, the synchronizing data in the $\sync$ message cannot be extracted. For the ranging request and response, since these signals are indistinguishable by the use of a secure PRF, the only information available to the adversary is the energy of the signals, which can be obtained by tracking the communication channel. We consider an adversary with a high-precision energy detector (e.g., using a high-quality low-noise amplifier and much higher sample rate than the rate used by ranging nodes). By tracking the energy of the request and response arriving at itself, the adversary can obtain the ranging session's start time $\hat{t}_S$ and end time $\hat{t}_R$ in its own time reference.
\begin{align}
\hat{t}_S & = t_S+ToF_{\initiator\rightarrow\adversary} \\
\hat{t}_R & =t_S+ToF+T_W+T_{\resp}+ToF_{\reflector\rightarrow\adversary}.
\end{align}
The ToF between the initiator and reflector can be written as
\begin{equation}
\label{eq:tof-adversary}
ToF = \hat{t}_R - \hat{t}_S - T_W - T_{\resp} - ToF_{\reflector\rightarrow\adversary} + ToF_{\initiator\rightarrow\adversary}
\end{equation}

It is seen from \Cref{eq:tof-adversary} that for a generic scenario, when the adversary is not aware of its distance to the initiator and reflector (i.e., unknown $ToF_{\initiator\rightarrow\adversary}$ and $ToF_{\reflector\rightarrow\adversary}$), the ToF cannot be reliably estimated by the adversary even if the response delay $T_W$ is known. Now consider special cases where $ToF_{\initiator\rightarrow\adversary},ToF_{\reflector\rightarrow\adversary}$ can be cancelled out, for instance the adversary is located in the initiator's proximity ($ToF_{\initiator\rightarrow\adversary}=0$ and $ToF_{\reflector\rightarrow\adversary}=ToF$) or right in the middle between initiator and reflector ($ToF_{\initiator\rightarrow\adversary}=ToF_{\reflector\rightarrow\adversary}$), the random waiting period $T_W$ would hinder the adversary from distance estimation with an acceptable accuracy. It is worth noting that in the above equations, we already ignore the unknown timing error $T_E$, which would even increase challenges to the adversary.
\section{System and Performance Evaluation}
\label{sec:evaluation}

In this section, we evaluate the performance of our ranging protocol.
Our evaluation metric is the distance error obtained from a ranging session, which is the difference between the ground truth and the estimated value.
We first evaluate the basic functionality of our ranging protocol in a real testbed over-the-air environment for a pair ranging, with one initiator and one reflector.
To evaluate full features of our protocol, we extend the number of reflectors to create broadcast ranging scenarios. These scenarios are evaluated in simulation due to limited hardware availability.

\subsection{Setup and Methodology}
Throughout the evaluation, we fix the epoch duration to $\Delta\epoch=\SI{1}{s}$. Unless otherwise stated, we enable the Successive Interference Cancellation (SIC) and the response batch mode with the batch size $|\batch|=10$ by default. In both the testbed experiments and simulations, the sample rate is set to $f_S=\SI{100}{MHz}$ (unless otherwise noted), while the actual signal bandwidth $B$ can vary from $\SI{100}{MHz}$ to $\SI{25}{MHz}$. The center frequency is fixed at $\SI{2.45}{GHz}$. We use a sequence length of $L=512$ for all scenarios. The sequence detection uses the ratio threshold $\alpha=50$ and $L_0=256$.

For real testbed experiments, the setup consists of two ranging nodes. Each  node is composed of a SDR device and a host machine to control it.
The SDR device is a USRP X310 equipped with a UBX-160 daughterboard~\cite{wiki:usrpX310}. The device is mounted with two omnidirectional \SI{2.4}{\GHz} antennas on the same daughterboard, one for receiving and one for transmitting signals.
The initiator is hosted on a HP-Z620 workstation with 2X Intel Xeon E5-2670 2.6GHz 16-Cores, 64GB RAM connected to USRP via a $\SI{10}{Gbps}$ SFP+ cable that is able to sustain a data stream at sample rate $\SI{100}{MHz}$. The reflector runs directly on the USRP device with a custom-modified FPGA for ranging purpose (See \Cref{appendix:fpga}).

For simulation, we set up a wideband wireless environment using Matlab's Wideband LOS Channel. This channel models the propagation of RF signals between multiple points with a Line-of-Sight in space and includes the free-space attenuation as well as the time delay effect. The model is configured with an operating frequency of \SI{2.45}{\GHz} and sample rate of \SI{100}{\MHz}.
Ranging request and response signals are generated as a sequence of configurable $L$ random bits passed to a BPSK modulator. The modulated signals are then fed to the channel simulator to undergo the effects of distance such as path loss and time delay.
\subsection{Pair Ranging}
\subsubsection{Accuracy}
%%% Outdoor 
The pair ranging evaluation is carried out in a real testbed. The initiator and the reflector are placed at increasing distances of each other. Both nodes are at the same height of \SI{1.2}{\m} and operate at sample rate of \SI{100}{MHz}. Each node is configured to transmit with \SI{20}{\dB} gain and receive with \SI{30}{\dB} gain. The batch mode is used with the batch size $|\batch|=10$. For each specific ground truth distance, we perform 10 ranging sessions and compute the average estimated distance and average distance error (mean of absolute error).
\begin{figure}[h]
    \centering
    \includegraphics[width=0.47\textwidth]{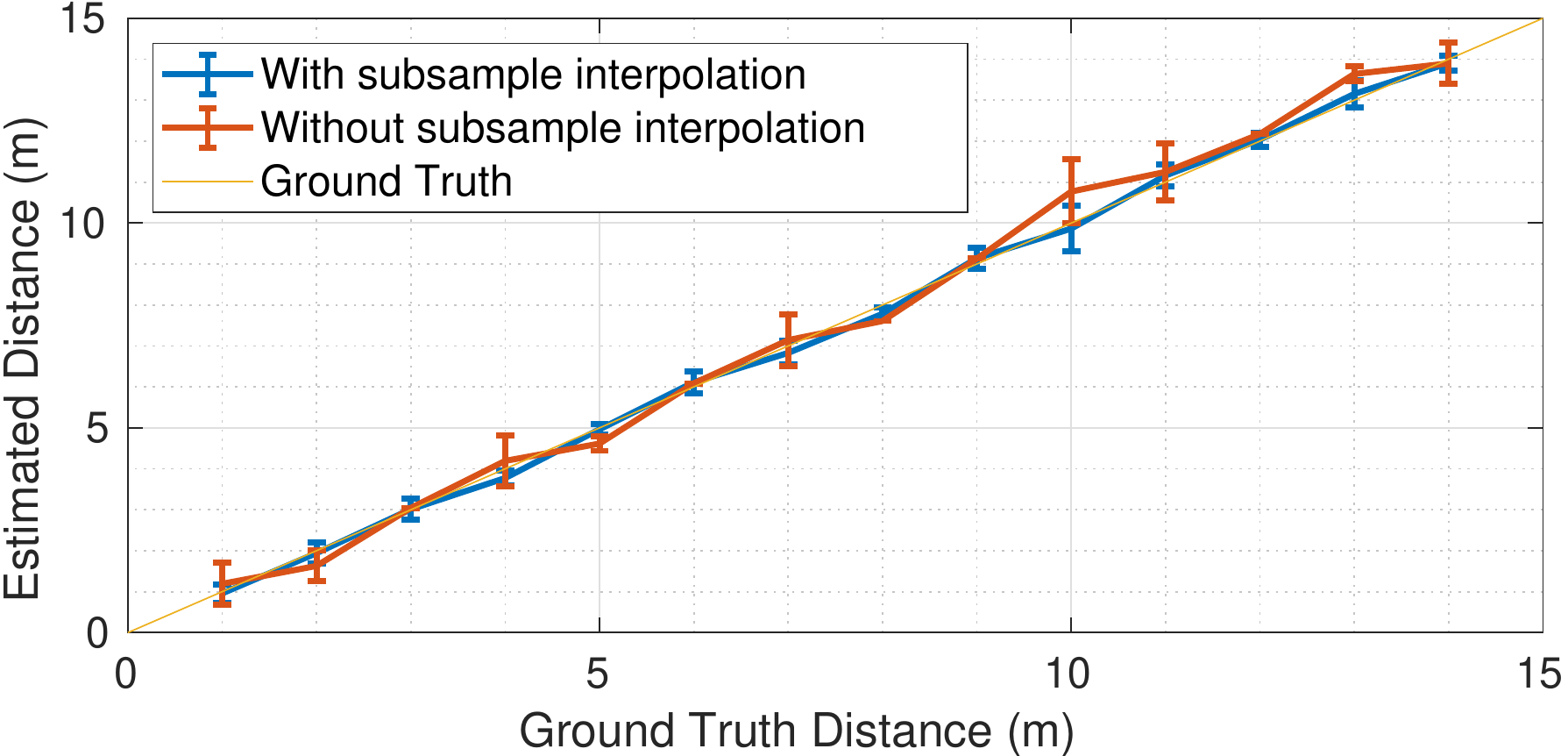}
    \caption{Pair ranging in outdoor environment.}
    \label{fig:outdoor}
\end{figure}
\begin{figure}[h]
    \centering
    \includegraphics[width=0.47\textwidth]{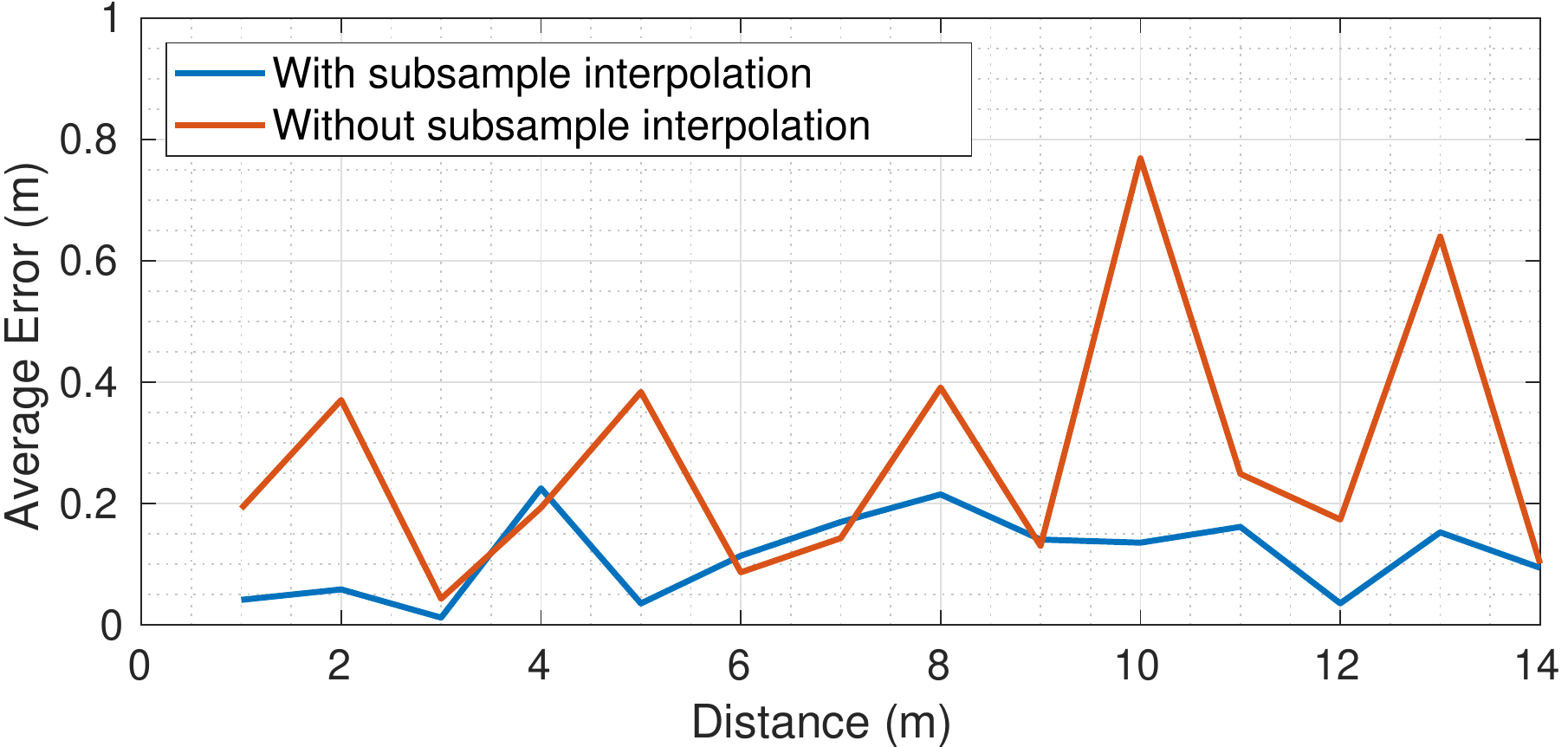}
    \caption{Accuracy of pair ranging in outdoor environment.}
    \label{fig:outdoor-error}
\end{figure}

\Cref{fig:outdoor,fig:outdoor-error} show the average estimated distances and average estimation error in an outdoor environment (a large backyard with fence and a few trees) for distances varying from \SI{1}{\m} to \SI{14}{\m}. To show the crucial role of timing error estimation with subsample interpolation, we also show the result of measurements without applying subsample interpolation. It can be clearly observed from \Cref{fig:outdoor-error} that the average distance error is around $\SI{15}{\cm}$ with subsample interpolation, while without this fine-grained timing estimation, the error is more fluctuating and can go up to \SI{75}{\cm}. In the latter case, the zig-zag pattern is the evidence for the coarse sampling resolution, where the error is high when the ToF is not a multiple of the sample period. The smoother curve with subsample interpolation proves its effectiveness in timing error correction.

%%% Indoor
\begin{figure}[h]
    \includegraphics[width=0.47\textwidth]{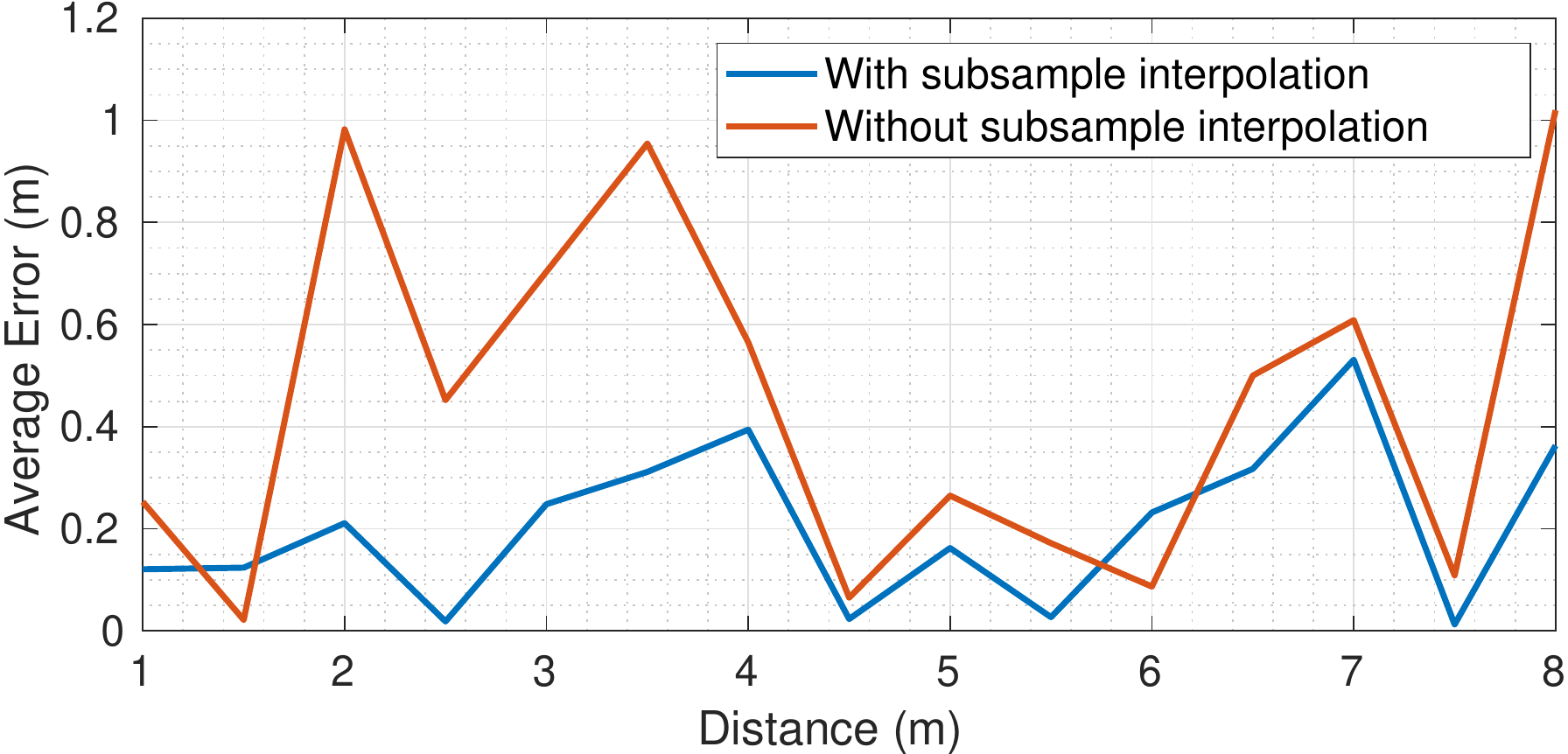}
    \caption{Accuracy of pair ranging in \textit{indoor} environment.}
    \label{fig:indoor-error}
\end{figure}
We also carry out an indoor experiment inside a $4\times9\times\SI{3.5}{\m}$ room with wood floor, doors and glass windows (\Cref{fig:indoor-error}). The estimated distances have an average error of \SI{20}{\cm}, which is slightly higher than the error in the outdoor experiment due to multipaths that degrade the estimation.
While the indoor accuracy depends on the specific environment, the results with subsample interpolation generally outperform those without this technique. The subsample estimation is also more stable due to the fine-grained resolution.

%%% SNR
\subsubsection{Robustness}
\begin{figure}[h]
    \centering
    \includegraphics[width=0.47\textwidth]{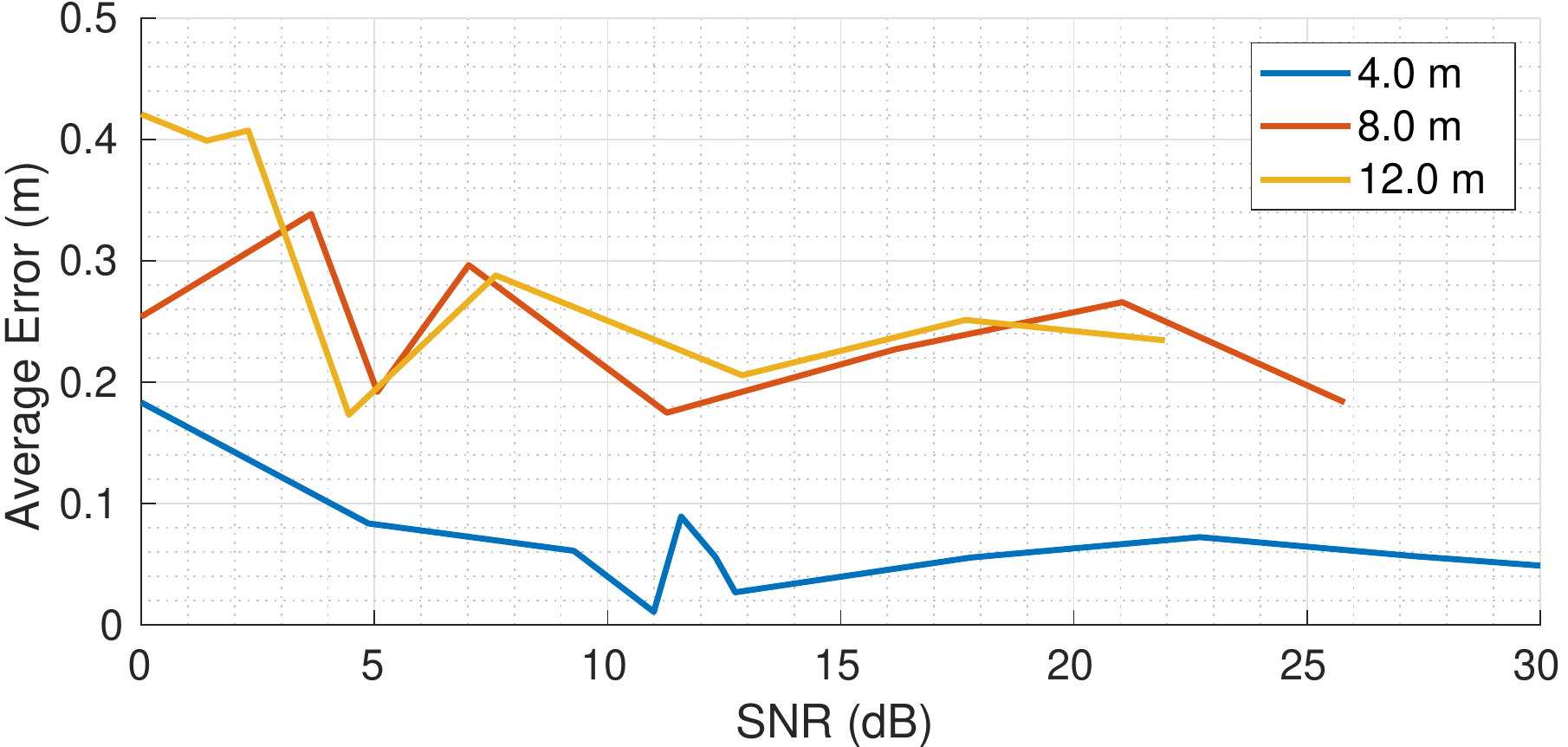}
    \caption{Accuracy under different SNR conditions for various distances in outdoor environment.}
    \label{fig:snr}
\end{figure}

We evaluate our system's robustness by conducting an experiment under different Signal-to-Noise Ratio (SNR) situations. There is no mutual interference in this pair ranging scenario. We compare the system in three different distance settings: $\SI{4}{m},\SI{8}{m},\SI{12}{m}$. For each run, we use a fixed receive gain while varying the transmit gain of both initiator and reflector to change the SNR level of the received signal at each node.
In \Cref{fig:snr}, the average distance error (over 10 runs) fluctuates with a deviation of \SI{\pm10}{\cm} over different SNRs. The result in this experiment is an indicator showing that our ranging protocol still achieves acceptable accuracy under low SNR regime.

%%% Upsample
\subsubsection{Spectrum Flexibility}
We evaluate the distance accuracy obtained in the upsampling mode, in which our ranging signals are shrunk to a narrower band. The original signals are upsampled with different interpolation factors  and low-pass filtered at a sample rate $f_S=\SI{100}{MHz}$ to generate signals of different bandwidths. For each bandwidth setting, we run the experiment to collect the estimations for a set of ground truth distances.
\begin{figure}[h]
    \centering
    \includegraphics[width=0.47\textwidth]{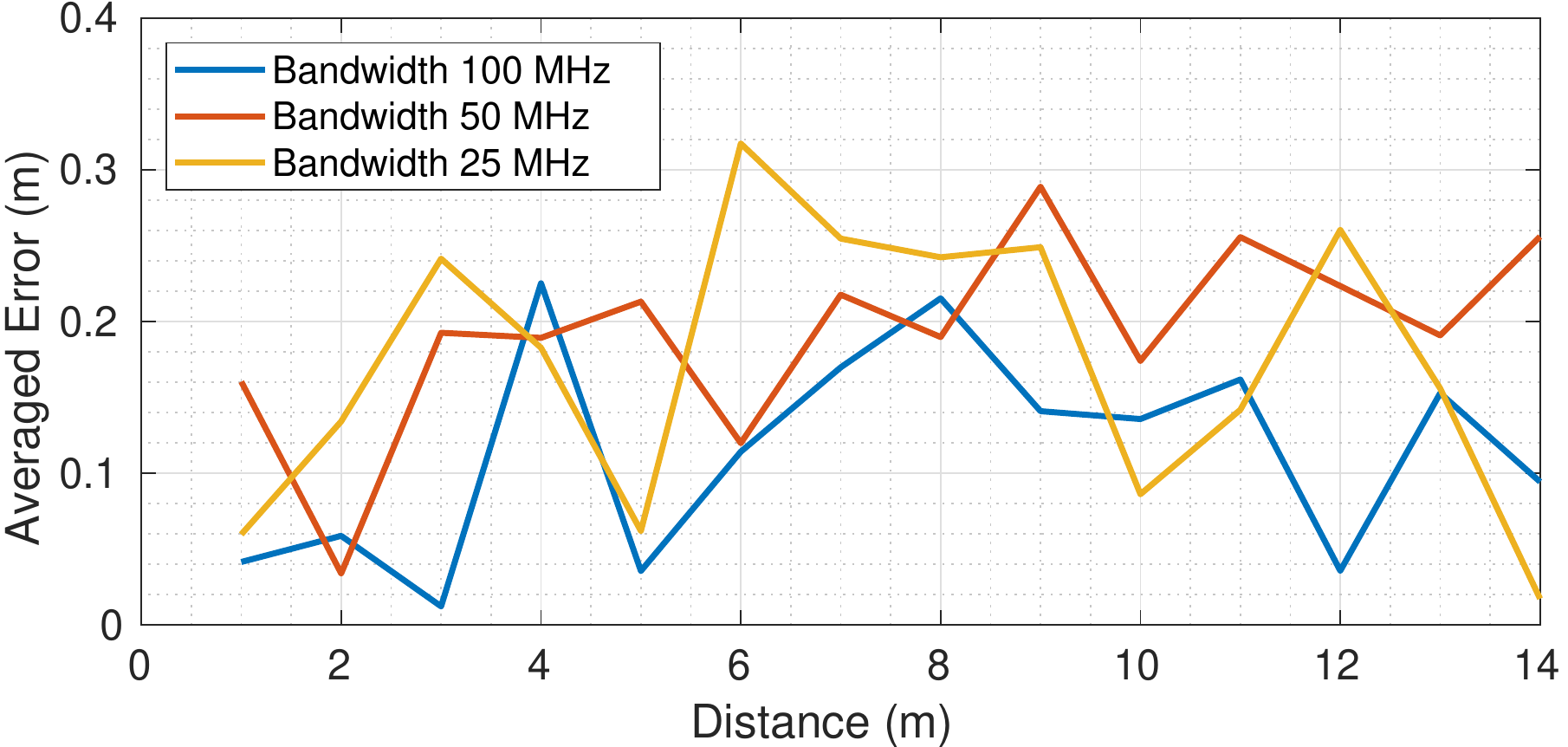}
    \caption{Outdoor pair ranging experiment with various signal bandwidths of \SI{100}{\MHz}, \SI{50}{\MHz}, \SI{25}{\MHz} using an upsampling factor of $1$, $2$, $4$ respectively. Sample rate is fixed at $\SI{100}{MHz}$.}
    \label{fig:upsample-outdoor-error}
\end{figure}

\Cref{fig:upsample-outdoor-error} shows the results of the upsampling experiment, where each data point is an average over 10 runs. The accuracy is comparable in all three cases with the average error of \SI{20}{\cm} on the evaluated distances. It is observed from this experiment that the distance accuracy in our system is not dependent on the ranging signal bandwidth. In fact, we can narrow down the ranging spectrum by 4 times without hurting the performance.

\subsubsection{Effect of Sample Rate}
As opposed to the signal bandwidth, the sample rate is the key factor impacting the ranging accuracy. To verify this theory, we conduct a separate experiment to analyze the impact of sample rate. In this experiment, we use the interpolation factor of 1, i.e., the signal bandwidth is always equal to the sample rate.
\begin{figure}[h]
    \centering
    \includegraphics[width=0.47\textwidth]{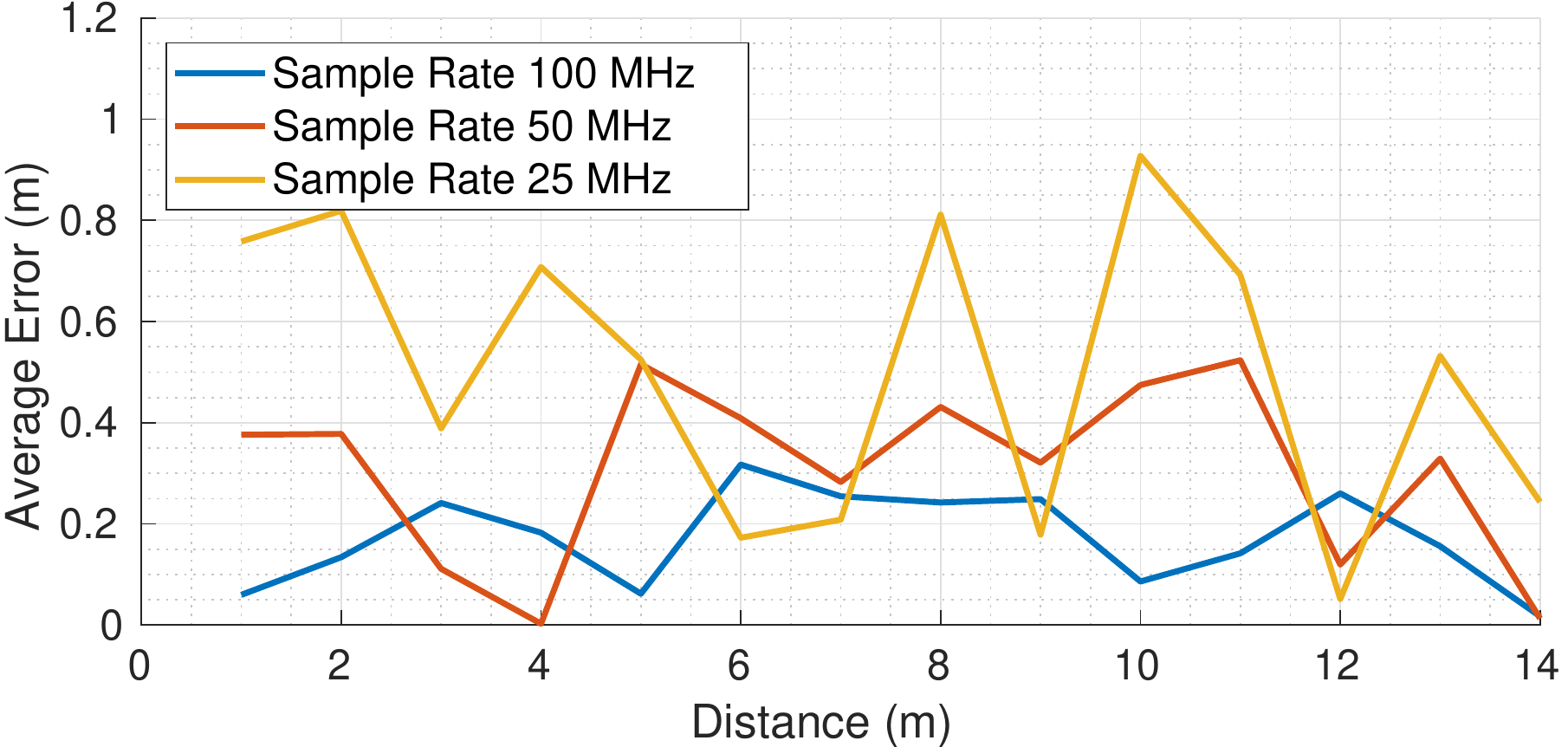}
    \caption{Accuracy impacted by different sample rates.}
    \label{fig:rate-outdoor-error}
\end{figure}
\Cref{fig:rate-outdoor-error} shows results of this experiment confirming the improvement in accuracy from \SI{50}{\cm} to \SI{30}{\cm} and \SI{15}{\cm} when the sample rate is increased from \SI{25}{MHz} to \SI{50}{MHz} and \SI{100}{MHz}, respectively.

\subsection{Broadcast Ranging}
In broadcast ranging, the key challenge is the mutual interference that can degrade the accuracy of distance estimation.
The SINR of a response coming from an individual reflector is reduced when more nodes participate in the ranging sessions.                         
The broadcast ranging evaluation in this section is carried out by simulation.

%%%%%%%%%%%%%%%%%%%%%%%%%%%%%%%%%%%%
\subsubsection{Equidistant Scenario}
We first evaluate the broadcast ranging by equidistant scenarios, where reflectors are located randomly on a circle with the same distance to the initiator. We focus on the effect of distance and number of reflectors on the system performance. 

\BfPara{Failure Rate}
As collisions in broadcast ranging can cause  nodes to disconnect, we  evaluate the system in terms of failure rate,  the ratio between the number of reflectors that fail the ranging operation (i.e., no distance is estimated by the initiator due to missed or corrupted messages) and the total number of reflectors in the system. Note that a reflector succeeds if at least one of its responses can reach the initiator.

\begin{figure}[h]
    \centering
    \includegraphics[width=0.47\textwidth]{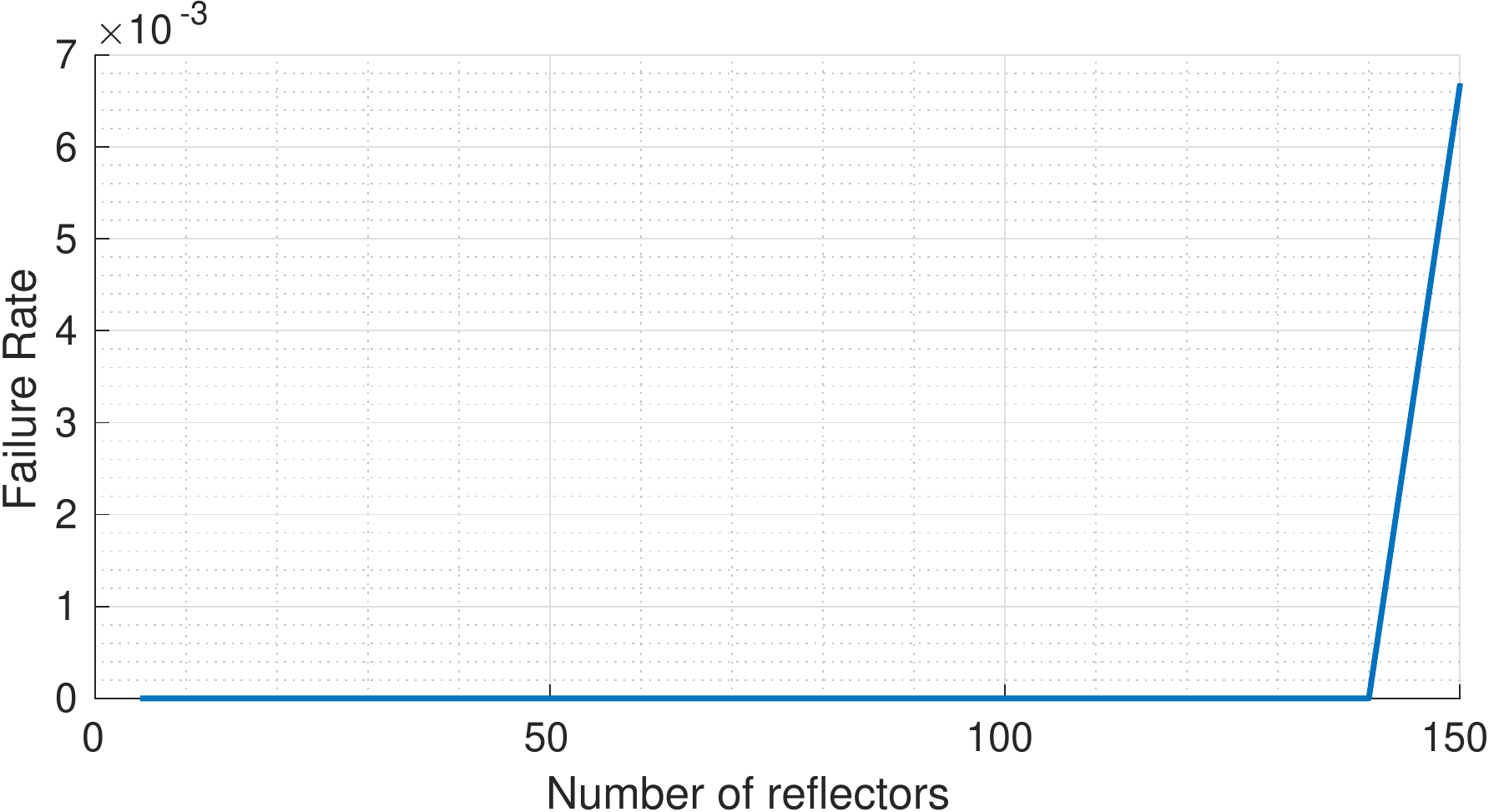}
    \caption{Failure rate of broadcast ranging with session duration of \SI{1}{\ms} in equidistant scenario simulation.}
    \label{fig:sim-equidistant-spread1ms-failure-rate}
\end{figure}

We recall that the random waiting period $T_W$ between responses is randomly selected within a waiting window $W$. Setting $W$ needs to take into account the trade-off between the collision rate and the ranging session's complete time. A large value of $W$ would reduce the broadcast collision at the expense of ranging time. In the following failure rate evaluation, we first choose $W$ such that the ranging session duration $T_{\mathrm{session}}$ is around $\SI{1}{ms}$. This implies ranging utilizes only $0.1\%$ of the epoch duration $\Delta\epoch=\SI{1}{s}$. We place the reflectors randomly on a circle with the same distance to the initiator. By changing the number of reflectors in the ranging session, we collect the failures for each run. \Cref{fig:sim-equidistant-spread1ms-failure-rate} shows that with $T_{\mathrm{session}}=\SI{1}{ms}$, there are almost no collisions for broadcast ranging up to 100 nodes. The failure rate slightly increases to $0.007$ when the number of reflectors reaches 150.

\begin{figure}[h]
    \centering
    \includegraphics[width=0.47\textwidth]{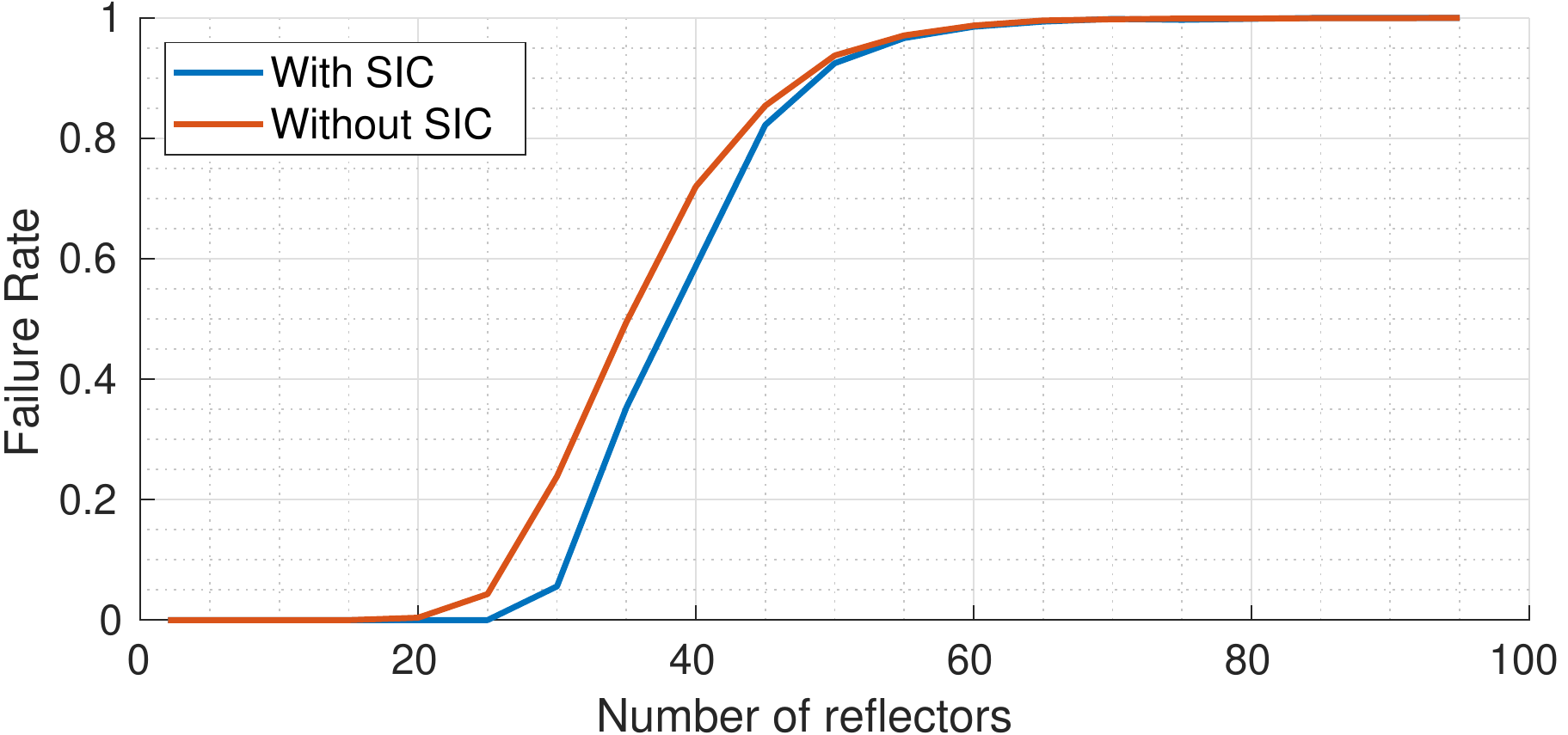}
    \caption{Failure rate of broadcast ranging with constrained session duration of $\SI{100}{us}$ in equidistant scenario simulation.}
    \label{fig:sim-equidistant-failure-rate}
\end{figure}

Now we are interested in the failure rate in a more challenging situation, where the session time is limited to $\SI{100}{us}$. With this constraint, the results in
\Cref{fig:sim-equidistant-failure-rate} show that system can sustain up $20$ reflectors, beyond that point more reflectors start to fail. However, if the system has only 20 or less reflectors, it would be advantageous to reduce the epoch duration to $\SI{100}{us}$ and the system can support up to 10,000 ranging sessions per second. From \Cref{fig:sim-equidistant-failure-rate}, we also see the benefit of SIC that slightly improves the failure rate. 

\begin{figure}[h]
    \centering
    \includegraphics[width=0.47\textwidth]{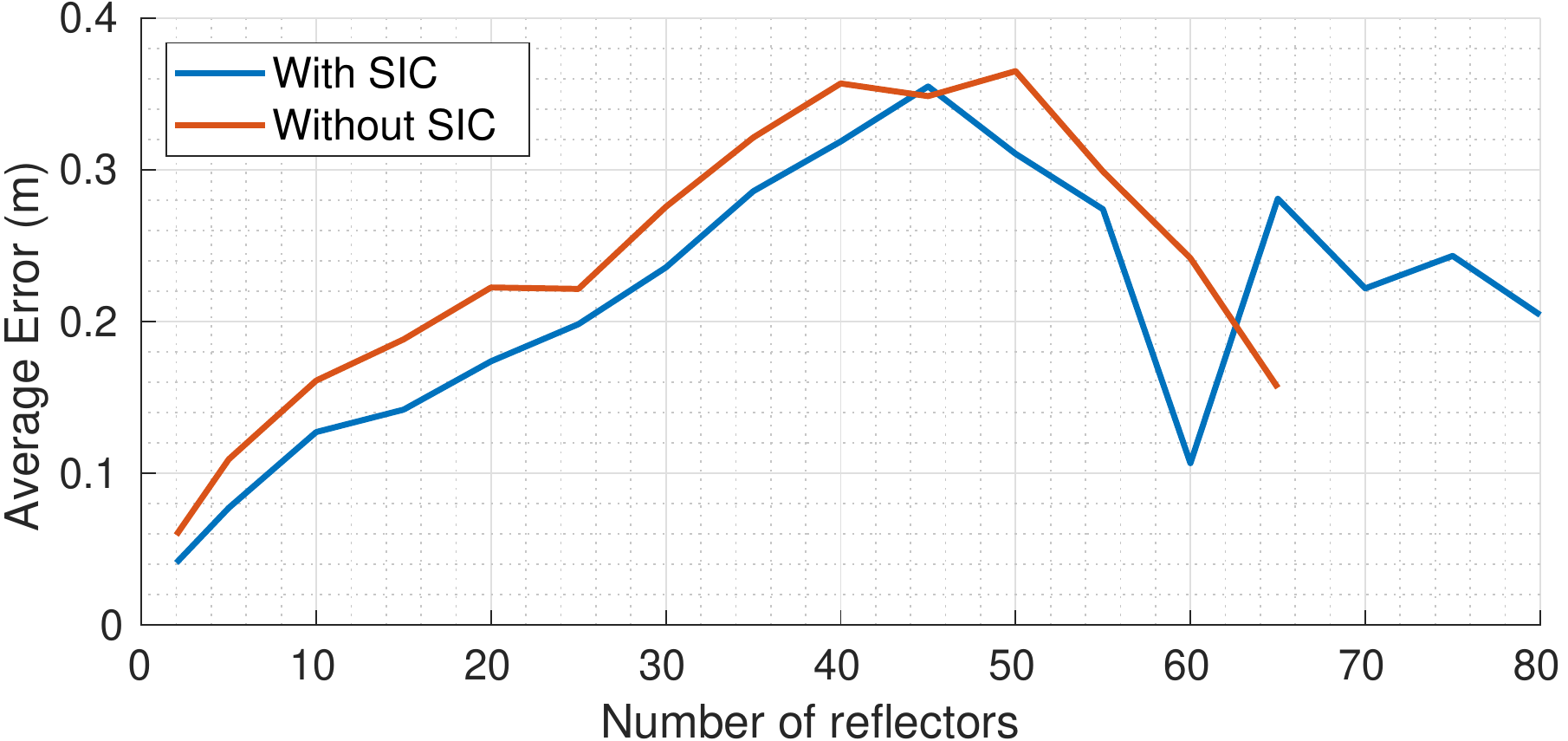}
    \caption{Impact of number of reflectors on broadcast ranging with constrained session duration of $\SI{100}{us}$ in equidistant simulation.}
    \label{fig:sim-equidistant-error-nnodes}
\end{figure}

\BfPara{Impact of Number of Nodes}
Now we evaluate how the number of ranging nodes in a session affects the average accuracy. Specifically, for each run, we select a fixed number of ranging nodes, then we increase the circle radius from $\SI{1}{m}$ to $\SI{25}{m}$ with reflectors randomly distributed on the circle. To fully understand the impact of collisions, we constrain all responses to $\SI{100}{us}$ (10 responses of $\SI{5}{us}$ per reflector using $\SI{100}{MHz}$ bandwidth providing 10,000 ranging epochs per second). The average distance error is collected based on all distance settings in this run. The result of one run is represented by a data point in~\Cref{fig:sim-equidistant-error-nnodes}. Recall that the Successive Interference Cancellation (SIC) is used to remove mutual interference from the combined received ranging signal at the initiator. To see the effectiveness of the SIC, we rerun the system on the same data with SIC disabled.

\Cref{fig:sim-equidistant-error-nnodes} shows that the overall error is well below \SI{40}{\cm} and the SIC can improve the accuracy by \SI{5}{\cm} in most scenarios.
Now focusing on the accuracy, we see that the accuracy degrades due to the overlapping of responses from multiple reflectors. Although our random sequences have good correlation property,  massive collisions can severely degrade the accuracy or even completely destroy the signals, which are indicated by the disappearance of the tail on the red curve (without SIC). 

An interesting part in~\Cref{fig:sim-equidistant-error-nnodes} is the dip in both curves when the number of reflectors increases beyond 50. When collisions grow over a threshold, many reflectors become disconnected while the remaining ones yield lower errors. Therefore, the average error, as computed on the responses that actually arrived, is lower but does not mean an improved performance. 

\begin{figure}[h]
    \centering
    \includegraphics[width=0.47\textwidth]{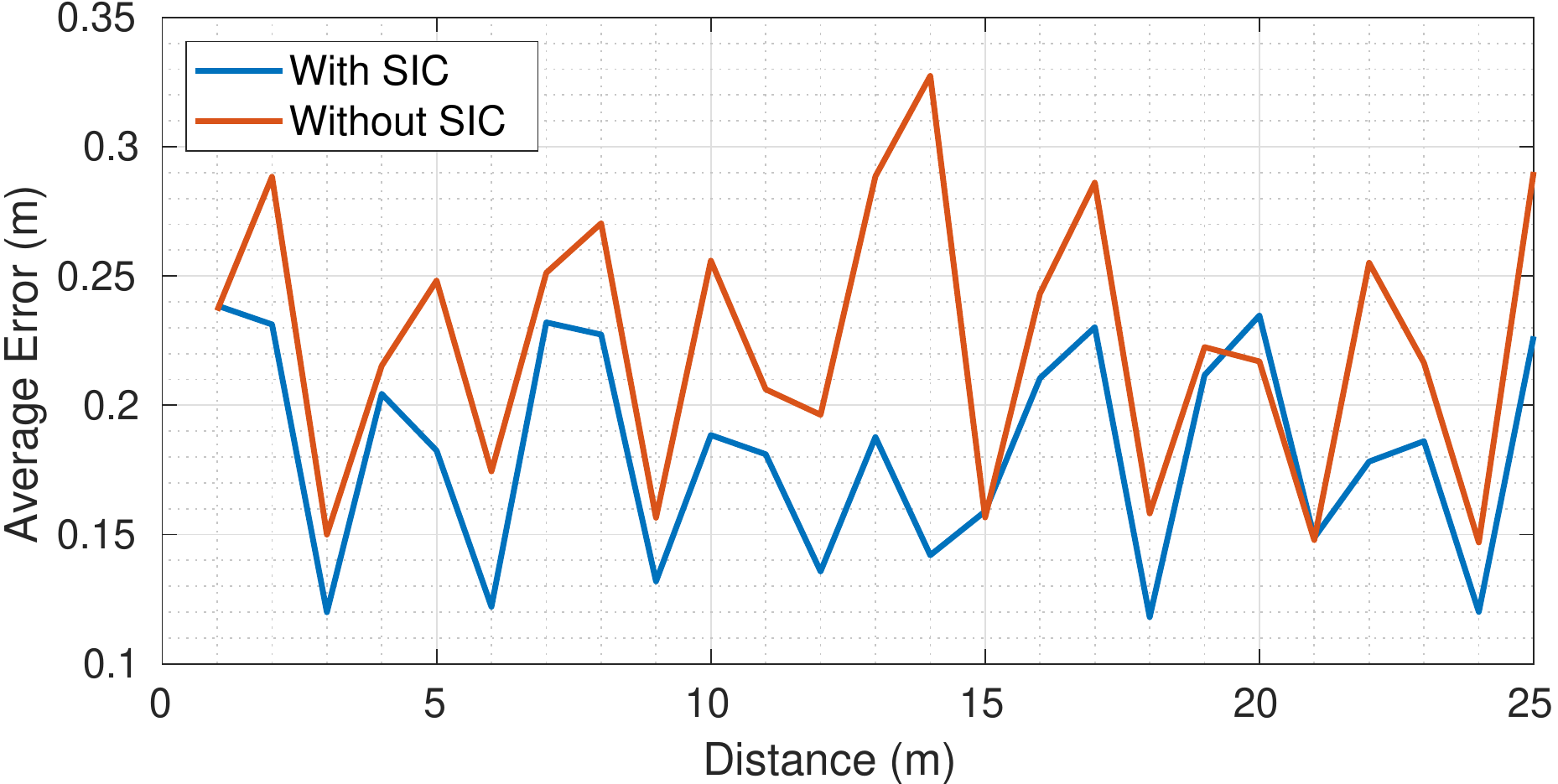}
    \caption{Impact of distance on broadcast ranging with constrained session duration of $\SI{100}{us}$ in equidistant scenario simulation.}
    \label{fig:sim-equidistant-error-distance}
\end{figure}

\BfPara{Impact of Distance}
We evaluate how  distance affects broadcast ranging. For each run, we select a new distance and change the number of ranging nodes. The average error is computed over all the sessions. 
\Cref{fig:sim-equidistant-error-distance} shows  how the error changes when increasing  distance from $\SI{1}{m}$ to $\SI{25}{m}$. In this experiment, we also observe a boost of \SI{5}{\cm} by applying the SIC technique. The average error fluctuates around \SI{15}{\cm}, but  is independent of the ranging distance. This result is aligned with the pair ranging case already seen in~\Cref{fig:outdoor-error}.

%%%%%%%%%%%%%%%%%%%%%%%%%%%%%%%%%%%%
\subsubsection{Random Scenario}
In a generic broadcast ranging scenario, the accuracy depends mainly on the number of reflectors and their position. Far-away reflectors typically give less accurate estimation due to weaker response signal, which additionally experiences strong interference from those reflectors who are closer to the initiator. To evaluate the system in such scenario, we have run 100 scenarios with different number of reflectors randomly located at different distances within a range of \SI{30}{\m} to the initiator. We constrain the session duration to roughly $\SI{100}{us}$ such that we can observe more collisions and evaluate the effectiveness of the SIC technique.

\begin{figure}[h]
    \centering
    \includegraphics[width=0.47\textwidth]{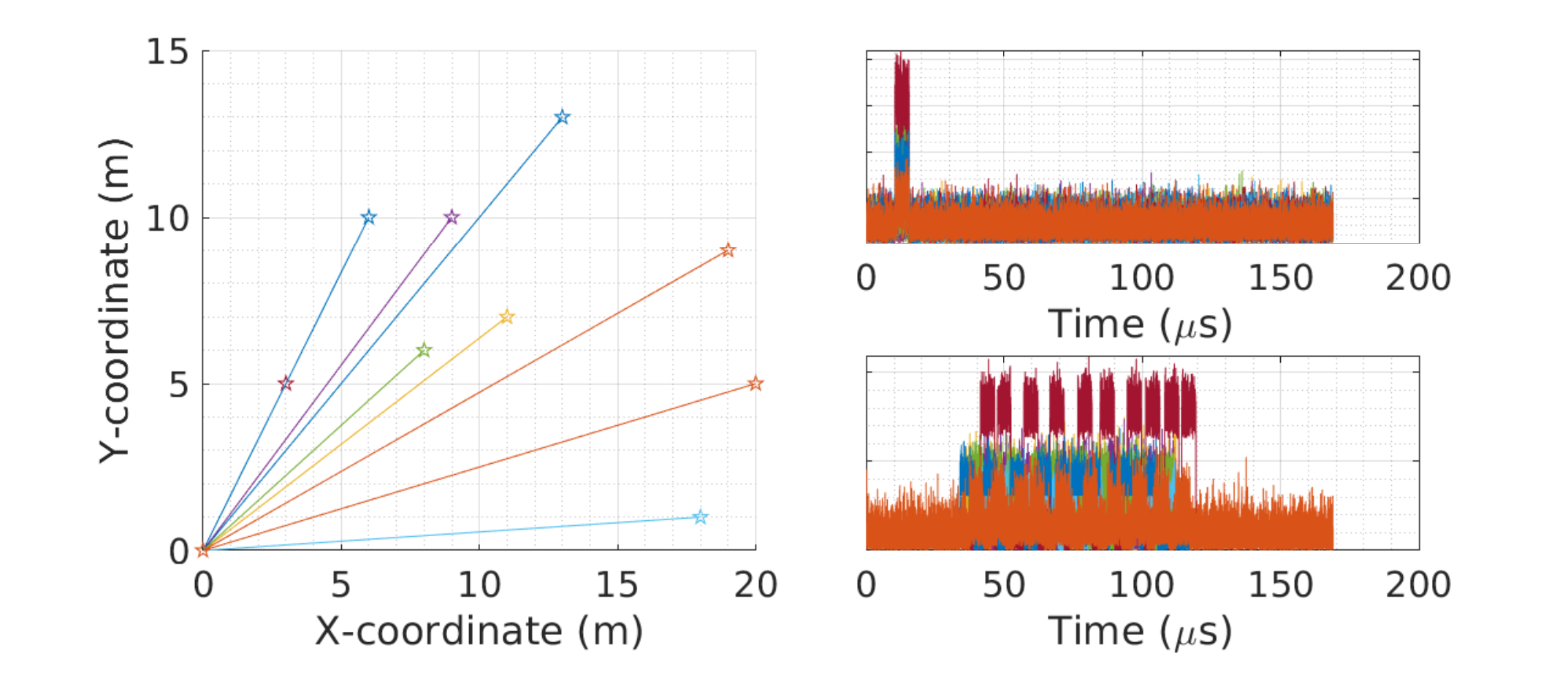}
    \caption{Scenario simulation of random positions for $9$ reflectors.}
    \label{fig:sim-random}
\end{figure}

\Cref{fig:sim-random} shows an example of a random scenario of $9$ reflectors.
The left side shows the distance of each reflector to the initiator (located at the origin). The right side shows how responses arrive and collide at the initiator.

\begin{figure}[h]
    \centering
    \includegraphics[width=0.47\textwidth]{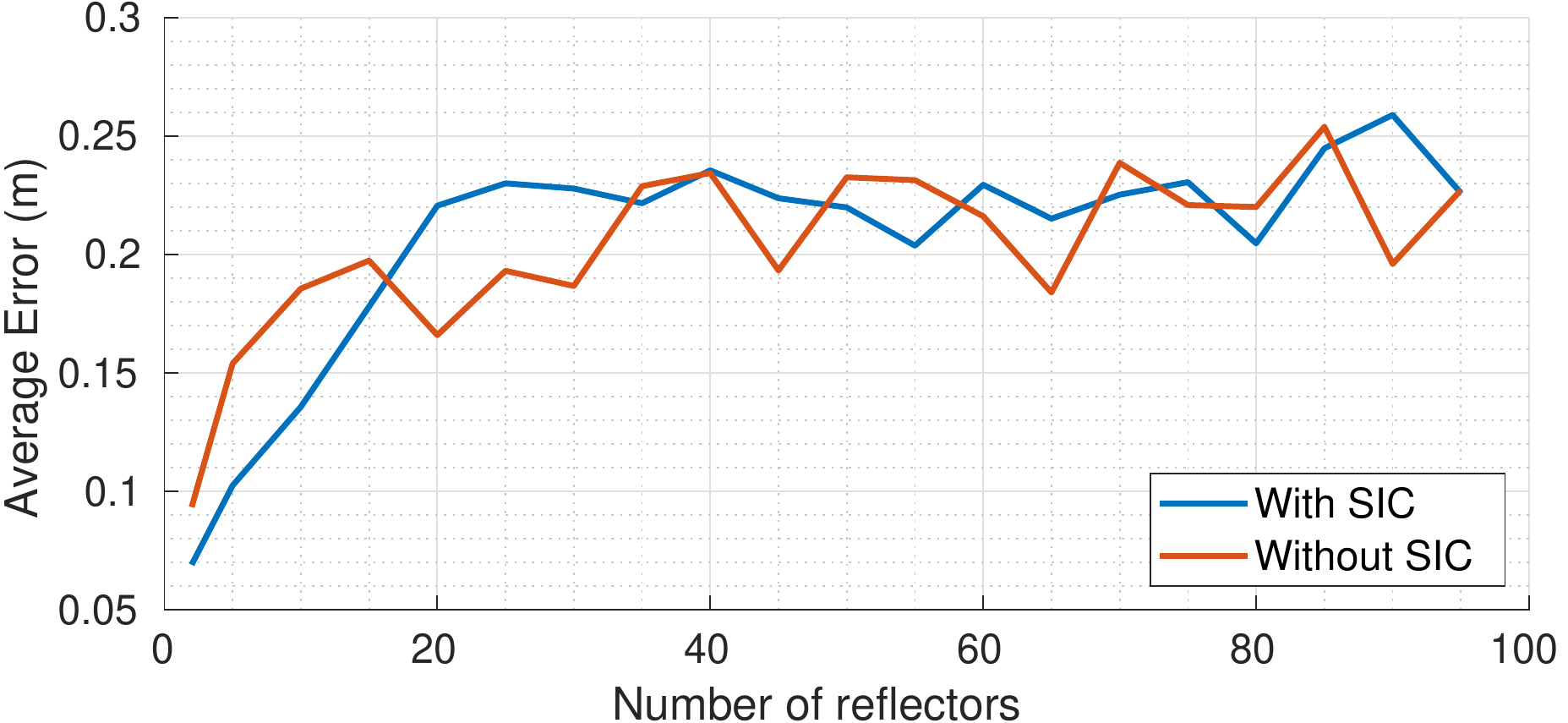}
    \caption{Impact of number of reflectors on broadcast ranging with session duration of $\SI{100}{us}$ in random scenario simulation.}
    \label{fig:sim-random-error-nnode}
\end{figure}
\begin{figure}[h]
    \centering
    \includegraphics[width=0.47\textwidth]{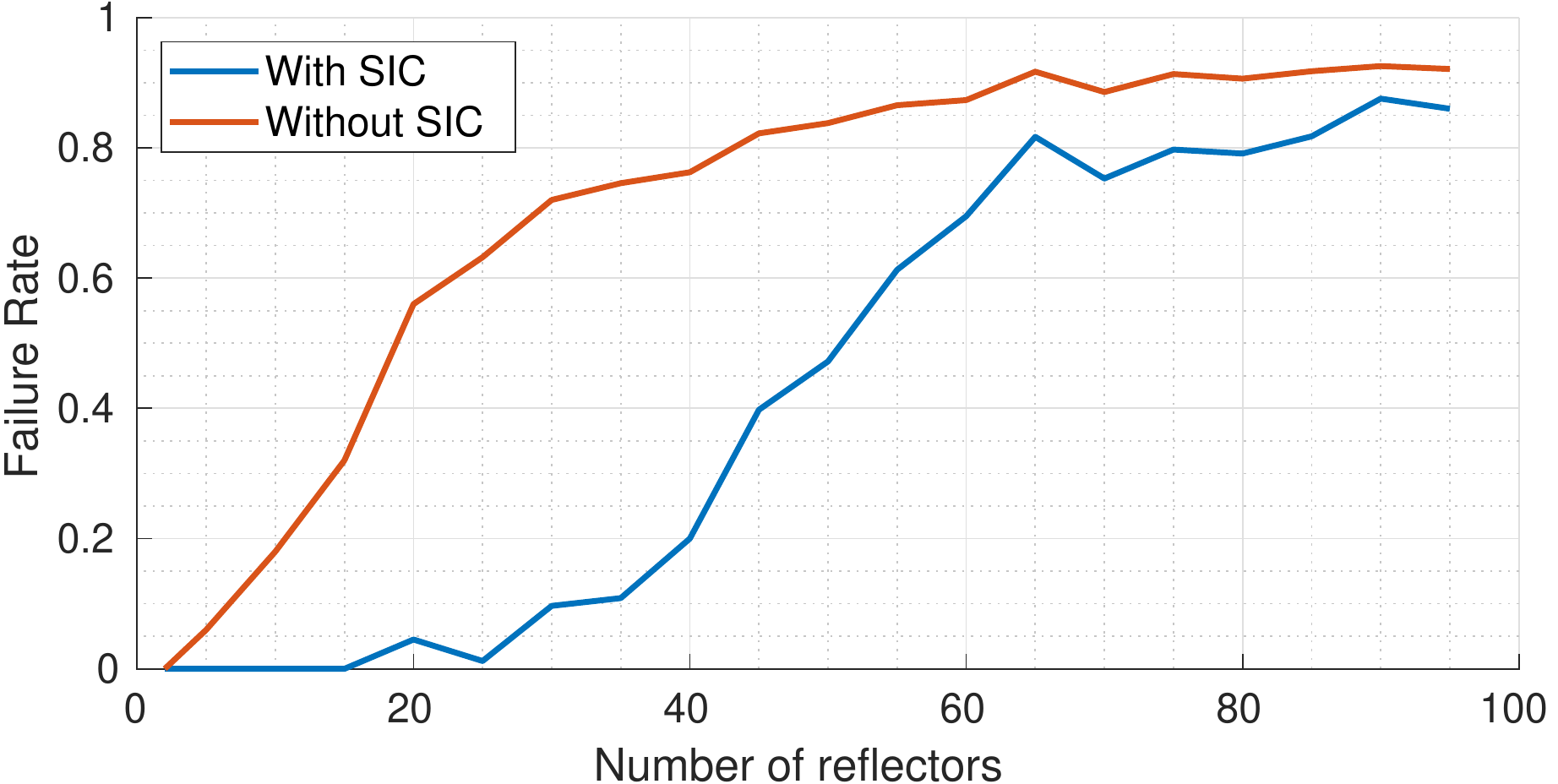}
    \caption{Failure rate of broadcast ranging with session duration of $\SI{100}{us}$ in random scenario simulation.}
    \label{fig:sim-random-failure-rate}
\end{figure}

The results for random scenarios are shown in \Cref{fig:sim-random-error-nnode,fig:sim-random-failure-rate}. The average error is stable around \SI{23}{\cm} when the number of reflectors exceeds $20$. From~\Cref{fig:sim-random-error-nnode}, the performance looks the same for both cases of enabling and disabling SIC. However, a closer look at~\Cref{fig:sim-random-failure-rate} reveals that without SIC there are more failures and only a small portion of reflectors succeeds, whose average estimation error is at an acceptable level. In fact, \Cref{fig:sim-random-failure-rate} shows a high failure rate of $0.57$ at $20$ reflectors when SIC is not applied, while the failure rate is below $0.1$ if SIC is used.
As a comparison to the equidistant scenarios (\Cref{fig:sim-equidistant-failure-rate}), the SIC clearly performs better under clear near-far effect when the random scenarios have more close-by and more far-away reflectors. 
\section{Conclusion}
We  proposed a secure broadcast ranging protocol with spectral flexibility that has minimal impact on accuracy. Flexibility is achieved through upsampling and successive interference cancellation. Stealth and security are achieved through cryptographically randomized messages in time and code.
We analyze its security over various attacks such as denial of service, distance enlargement, distance reduction and sniffing.
The protocol is designed for flexible implementation on FPGA and/or a host (with minimal FPGA modifications).
We evaluate our ranging system extensively under scenarios both in real over-the-air experiments for a pair of devices, and simulations for scalability to a large number of reflectors. We demonstrate through extensive performance evaluation that we can achieve an accuracy below \SI{20}{\cm} on a wide range of SNR (as low as $\SI{0}{dB}$), spectrum $\SI{25}{MHz}-\SI{100}{MHz}$ even when 20 simultaneous reflectors are constrained to sessions of $\SI{100}{us}$ leading to 10,000 simultaneous sessions per second. For sessions of $\SI{1}{ms}$ leading to 1,000 simultaneous sessions per second, the systems easily scales to over 100 reflectors.
% \newpage
% \input{appendix}

% using natbib
\bibliographystyle{abbrv}
\begin{spacing}{0.9}
    \setlength{\bibsep}{6pt}
    \bibliography{paper}

\begin{thebibliography}{10}

\bibitem{3dbaccess}
3db Access.
\newblock 3db6830 datasheet.
\newblock \url{http://www.3db-access.com/technology}, 2020.

\bibitem{wiki:uhd}
E.~K. Base.
\newblock Uhd --- ettus knowledge base{,}.
\newblock \url{https://kb.ettus.com/index.php?title=UHD&oldid=4720}, 2020.
\newblock [Online; accessed 11-May-2020].

\bibitem{wiki:usrpX310}
E.~K. Base.
\newblock X300/x310 --- ettus knowledge base{,}.
\newblock \url{https://kb.ettus.com/index.php?title=X300/X310&oldid=4930},
  2020.
\newblock [Online; accessed 11-May-2020].

\bibitem{brandchaum01db}
S.~Brands and D.~Chaum.
\newblock Distance-bounding protocols.
\newblock In T.~Helleseth, editor, {\em Advances in Cryptology --- EUROCRYPT
  '93}, pages 344--359, Berlin, Heidelberg, 1994. Springer Berlin Heidelberg.

\bibitem{brelurut}
A.~Brelurut, D.~Gerault, and P.~Lafourcade.
\newblock Survey of distance bounding protocols and threats.
\newblock In J.~Garcia-Alfaro, E.~Kranakis, and G.~Bonfante, editors, {\em
  Foundations and Practice of Security}, pages 29--49, Cham, 2016. Springer
  International Publishing.

\bibitem{CostinA2012}
A.~Costin and A.~Francillon.
\newblock {Ghost in the Air (Traffic): On insecurity of ADS-B protocol and
  practical attacks on ADS-B devices}.
\newblock {\em BlackHat USA 2012}, 2012.

\bibitem{dardari09}
D.~{Dardari}, A.~{Conti}, U.~{Ferner}, A.~{Giorgetti}, and M.~Z. {Win}.
\newblock Ranging with ultrawide bandwidth signals in multipath environments.
\newblock {\em Proceedings of the IEEE}, 97(2):404--426, 2009.

\bibitem{decawave}
Decawave.
\newblock Dw1000 user manual.
\newblock \url{https://www.decawave.com/dw1000/usermanual/}, 2020.
\newblock [Online; accessed 11-May-2020].

\bibitem{manuel10dd}
M.~Flury, M.~Poturalski, P.~Papadimitratos, J.-P. Hubaux, and J.-Y. Le~Boudec.
\newblock Effectiveness of distance-decreasing attacks against impulse radio
  ranging.
\newblock In {\em Proceedings of the Third ACM Conference on Wireless Network
  Security}, WiSec ’10, page 117–128, New York, NY, USA, 2010. Association
  for Computing Machinery.

\bibitem{Francillon2010RelayAO}
A.~Francillon, B.~Danev, and S.~Capkun.
\newblock Relay attacks on passive keyless entry and start systems in modern
  cars.
\newblock {\em IACR Cryptology ePrint Archive}, 2010:332, 2010.

\bibitem{url:gnu-radio}
{GNU Radio Website}.
\newblock \url{http://www.gnuradio.org}, 2020.
\newblock [Online; accessed 11-May-2020].

\bibitem{gps-sdr-sim}
{GPS SDR SIM}.
\newblock \url{https://github.com/osqzss/gps-sdr-sim}.
\newblock [Online; accessed 28-Aug-2020].

\bibitem{hancke08rfid}
G.~P. Hancke and M.~G. Kuhn.
\newblock Attacks on time-of-flight distance bounding channels.
\newblock In {\em Proceedings of the First ACM Conference on Wireless Network
  Security}, WiSec ’08, page 194–202, New York, NY, USA, 2008. Association
  for Computing Machinery.

\bibitem{ieee802154}
IEEE.
\newblock Ieee standard for low-rate wireless networks.
\newblock {\em IEEE Std 802.15.4-2015 (Revision of IEEE Std 802.15.4-2011)},
  pages 1--709, 2016.

\bibitem{manikanta15spotfi}
M.~Kotaru, K.~Joshi, D.~Bharadia, and S.~Katti.
\newblock Spotfi: Decimeter level localization using wifi.
\newblock In {\em Proceedings of the 2015 ACM Conference on Special Interest
  Group on Data Communication}, SIGCOMM ’15, page 269–282, New York, NY,
  USA, 2015. Association for Computing Machinery.

\bibitem{leu20mtac}
P.~Leu, M.~Singh, M.~Roeschlin, K.~G. Paterson, and S.~Capkun.
\newblock Message time of arrival codes: A fundamental primitive for secure
  distance measurement.
\newblock In {\em 2020 IEEE Symposium on Security and Privacy (SP)}, pages
  500--516, Los Alamitos, CA, USA, may 2020. IEEE Computer Society.

\bibitem{NarainRN2019}
S.~Narain, A.~Ranganathan, and G.~Noubir.
\newblock Security of {GPS/INS} based on-road location tracking systems.
\newblock In {\em Proceedings of the IEEE Symposium on Security and Privacy
  (SP)}, 2019.

\bibitem{gps-spoofing-maritime}
{New Scientist}.
\newblock Mass gps spoofing attack in black sea?
\newblock
  \url{https://maritime-executive.com/editorials/mass-gps-spoofing-attack-in-black-sea},
  July 2017.
\newblock [Online; accessed 28-Aug-2020].

\bibitem{gps-spoofing-new-scientist}
{New Scientist}.
\newblock Ships fooled in gps spoofing attack suggest russian cyberweapon.
\newblock
  \url{https://www.newscientist.com/article/2143499-ships-fooled-in-gps-spoofing-attack-suggest-russian-cyberweapon/#},
  August 2017.
\newblock [Online; accessed 28-Aug-2020].

\bibitem{NohKSSKCK2019}
J.~Noh, Y.~Kwon, Y.~Son, H.~Shin, D.~Kim, J.~Choi, and Y.~Kim.
\newblock Tractor beam: Safe-hijacking of consumer drones with adaptive gps
  spoofing.
\newblock {\em ACM Transactions on Privacy and Security}, 22(2), Apr. 2019.

\bibitem{OlafsdottirRC2017}
H.~{\'O}lafsd{\'o}ttir, A.~Ranganathan, and S.~Capkun.
\newblock On the security of carrier phase-based ranging.
\newblock In W.~Fischer and N.~Homma, editors, {\em Cryptographic Hardware and
  Embedded Systems -- CHES 2017}, pages 490--509, Cham, 2017. Springer
  International Publishing.

\bibitem{olafsdottir2017security}
H.~{\'O}lafsd{\'o}ttir, A.~Ranganathan, and S.~Capkun.
\newblock On the security of carrier phase-based ranging.
\newblock {\em Cryptographic Hardware and Embedded Systems--CHES 2017},
  10529:490--509, 2017.

\bibitem{poturalski10cicada}
M.~{Poturalski}, M.~{Flury}, P.~{Papadimitratos}, J.~{Hubaux}, and J.~{Le
  Boudec}.
\newblock The cicada attack: Degradation and denial of service in ir ranging.
\newblock In {\em 2010 IEEE International Conference on Ultra-Wideband},
  volume~2, pages 1--4, 2010.

\bibitem{poturalski11db}
M.~{Poturalski}, M.~{Flury}, P.~{Papadimitratos}, J.~{Hubaux}, and J.~{Le
  Boudec}.
\newblock Distance bounding with ieee 802.15.4a: Attacks and countermeasures.
\newblock {\em IEEE Transactions on Wireless Communications}, 10(4):1334--1344,
  2011.

\bibitem{RanganathanC2017}
A.~{Ranganathan} and S.~{Capkun}.
\newblock Are we really close? verifying proximity in wireless systems.
\newblock {\em IEEE Security Privacy}, 15(3):52--58, 2017.

\bibitem{aanjhan12cbattack}
A.~Ranganathan, B.~Danev, A.~Francillon, and S.~Capkun.
\newblock Physical-layer attacks on chirp-based ranging systems.
\newblock In {\em Proceedings of the Fifth ACM Conference on Security and
  Privacy in Wireless and Mobile Networks}, WISEC ’12, page 15–26, New
  York, NY, USA, 2012. Association for Computing Machinery.

\bibitem{rasmussen07secnav}
K.~B. Rasmussen, S.~Capkun, and M.~Cagalj.
\newblock Secnav: Secure broadcast localization and time synchronization in
  wireless networks.
\newblock In {\em Proceedings of the 13th Annual ACM International Conference
  on Mobile Computing and Networking}, MobiCom ’07, page 310–313, New York,
  NY, USA, 2007. Association for Computing Machinery.

\bibitem{rasmussen08db}
K.~B. Rasmussen and S.~\v{C}apkun.
\newblock Location privacy of distance bounding protocols.
\newblock In {\em Proceedings of the 15th ACM Conference on Computer and
  Communications Security}, CCS ’08, page 149–160, New York, NY, USA, 2008.
  Association for Computing Machinery.

\bibitem{Rogaway2004}
P.~Rogaway.
\newblock Nonce-based symmetric encryption.
\newblock In {\em International Workshop on Fast Software Encryption}, pages
  348--358. Springer, 2004.

\bibitem{sark17usrpranging}
V.~{Sark}, N.~{Maletic}, J.~{Gutiérrez}, and E.~{Grass}.
\newblock An approach for implementation of ranging and positioning methods on
  a software defined radio.
\newblock In {\em 2017 14th Workshop on Positioning, Navigation and
  Communications (WPNC)}, pages 1--6, 2017.

\bibitem{SathayeSRN2019}
H.~Sathaye, D.~Schepers, A.~Ranganathan, and G.~Noubir.
\newblock Wireless attacks on aircraft instrument landing systems.
\newblock In {\em 28th {USENIX} Security Symposium, {USENIX} Security 2019,
  Santa Clara, CA, USA, August 14-16, 2019.}, pages 357--372, 2019.

\bibitem{singh19uwbed}
M.~Singh, P.~Leu, A.~Abdou, and S.~Capkun.
\newblock Uwb-ed: Distance enlargement attack detection in ultra-wideband.
\newblock In {\em 28th {USENIX} Security Symposium ({USENIX} Security 19)},
  pages 73--88, Santa Clara, CA, Aug. 2019. {USENIX} Association.

\bibitem{singh19uwbpr}
M.~Singh, P.~Leu, and S.~Capkun.
\newblock Uwb with pulse reordering:securing ranging against relay and
  physical-layer attacks.
\newblock In {\em Proceedings of the 26th Annual Network and Distributed System
  Security Symposium}, page 06B2, Reston, VA, 2019. Internet Society.
\newblock 26th Annual Network and Distributed System Security Symposium (NDSS
  2019); Conference Location: San Diego, CA, USA; Conference Date: February
  24-27, 2019.

\bibitem{StrohmeierSPLM2017}
M.~Strohmeier, M.~Sch{\"a}fer, R.~Pinheiro, V.~Lenders, and I.~Martinovic.
\newblock On perception and reality in wireless air traffic communication
  security.
\newblock {\em IEEE Transactions on Intelligent Transportation Systems}, 2017.

\bibitem{svilainis2008analysis}
L.~Svilainis and V.~Dumbrava.
\newblock Analysis of the interpolation techniques for time-of-flight
  estimation.
\newblock {\em Ultragarsas" Ultrasound"}, 63(4):25--29, 2008.

\bibitem{svilainis13subsample}
L.~Svilainis, K.~Lukoseviciute, V.~Dumbrava, and A.~Chaziachmetovas.
\newblock Subsample interpolation bias error in time of flight estimation by
  direct correlation in digital domain.
\newblock {\em Measurement}, 46(10):3950 -- 3958, 2013.

\bibitem{taponecco13}
L.~{Taponecco}, P.~{Perazzo}, A.~A. {D'Amico}, and G.~{Dini}.
\newblock On the feasibility of overshadow enlargement attack on ieee 802.15.4a
  distance bounding.
\newblock {\em IEEE Communications Letters}, 18(2):257--260, 2014.

\bibitem{TippenhauerPRC2011}
N.~O. Tippenhauer, C.~P\"{o}pper, K.~B. Rasmussen, and S.~Capkun.
\newblock On the requirements for successful gps spoofing attacks.
\newblock In {\em Proceedings of the 18th ACM Conference on Computer and
  Communications Security}, CCS'11, 2011.

\bibitem{tippenhauer09temp}
N.~O. Tippenhauer, K.~B. Rasmussen, and S.~\v{C}apkun.
\newblock Secure ranging with message temporal integrity.
\newblock In {\em Cryptology ePrint Archive: Report 2009/602}, 2009.

\bibitem{vasisht16chronos}
D.~Vasisht, S.~Kumar, and D.~Katabi.
\newblock Decimeter-level localization with a single wifi access point.
\newblock In {\em 13th {USENIX} Symposium on Networked Systems Design and
  Implementation ({NSDI} 16)}, pages 165--178, Santa Clara, CA, Mar. 2016.
  {USENIX} Association.

\bibitem{viola2005spline}
F.~{Viola} and W.~F. {Walker}.
\newblock A spline-based algorithm for continuous time-delay estimation using
  sampled data.
\newblock {\em IEEE Transactions on Ultrasonics, Ferroelectrics, and Frequency
  Control}, 52(1):80--93, Jan 2005.

\bibitem{zafari2017survey}
F.~Zafari, A.~Gkelias, and K.~Leung.
\newblock A survey of indoor localization systems and technologies, 2017.

\bibitem{zanella16rssranging}
A.~{Zanella}.
\newblock Best practice in rss measurements and ranging.
\newblock {\em IEEE Communications Surveys Tutorials}, 18(4):2662--2686, 2016.

\end{thebibliography}
\end{spacing}

\appendix

\subsection{System Implementation}
We implement the secure broadcast protocol using GNU Radio platform and UHD framework.
In a ranging session, transmissions between nodes are BPSK-modulated signal from random binary sequences of $L$ bits. When the upsampling mode is enabled, the signal is interpolated with low-pass filter before being transmitted. Since random sequences are derived from the epoch $\epoch$, they can be pre-generated as soon as the epoch $\epoch$ is determined to reduce the processing delay.
On the receiving side, a signal detector is constructed for each sequence.
The cross-correlation is computed by an FIR filter with taps as the detecting sequence in the reversed order. The filter operation is optimized based on the FFT version of the FIR filter.

\BfPara{Initiator Implementation}
Regarding the initiator, who uses multiple detectors to simultaneously search for response sequences from multiple reflectors, high processing capability is required for heavy operations such as correlation computation and SIC procedure. For this reason, we implement the initiator on the host. 

\BfPara{Reflector on FPGA}
\label{appendix:fpga}
The low processing demand allows us to move crucial processing tasks of the reflector to FPGA.
The FPGA implementation removes the need to stream data from the SDR device to host for any signal processing task. By avoiding data streaming, we reduce a significant amount of communication and processing overhead for the host. This enables us to deploy a reflector even on a slow laptop, which not only provides mobility convenience, but also saves us significant time during over-the-air experiments.
Moreover, since data is processed directly in the FPGA, the operation follows a synchronous routine with respect to the master clock \SI{200}{\MHz}, and together with the capability to buffer responses in the memory, the FPGA provides a fast detection and reaction within couples of $\SI{5}{ns}$ clock cycles.

\BfPara{Hardware Latency}
In the host and FPGA implementation of initiator and reflector, there are processing latencies caused by the ADC (receive chain) and DAC (transmit chain) during the signal conversion between analog and digital domain. Therefore, the ToF expression in~\Cref{eq:timing1,eq:timing2} for the over-the-air experiments also include an additional hardware latency $T_{HW}$. The ToF with hardware latency would be rewritten as
\begin{equation*}
\widetilde{ToF} = \frac{1}{2}(t_R-t_S-T_W-T_{\var{\resp}}-T_{HW}) - T_E
\end{equation*}
Since $T_{HW}$ is constant, it can be determined through a linear fitting on experimental values. We perform extensive measurements between two devices at difference distances. The data is collected for each distance in range from \SI{1}{\m} to \SI{14}{\m} with steps of \SI{1}{\m} in a large backyard with a line of sight. At each position, we perform $1000$ measurements and derive the result.
We verified that this hardware latency value is applicable to devices of the same model.
\subsection{Secure Broadcast Ranging Protocol}
\begin{figure*}[t]
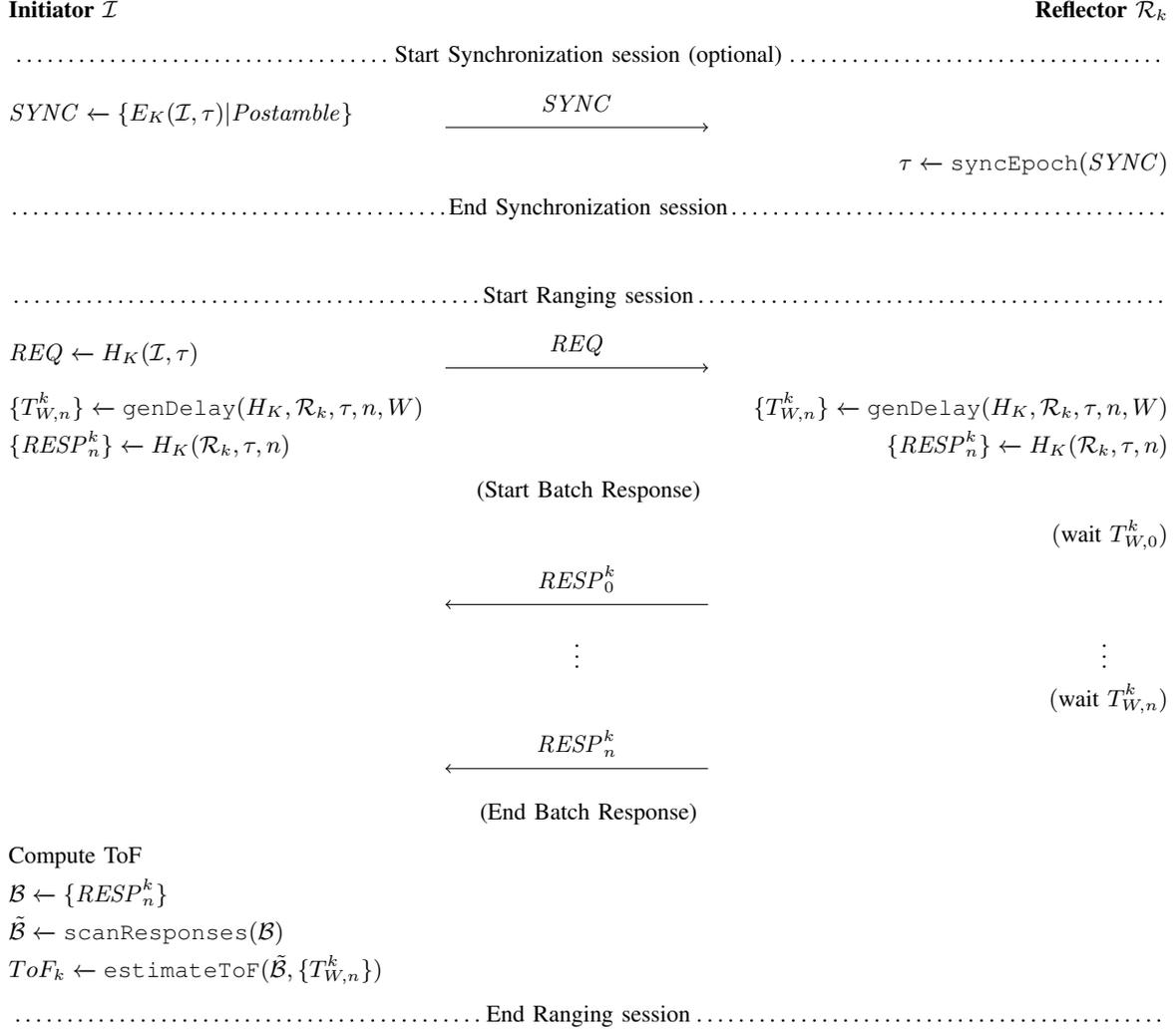

\procedureblock{}{
    \textbf{Initiator}~\initiator \>    \< \textbf{Reflector}~\reflector_k
    %% synchronization
    \pclb \pcintertext[dotted]{Start Synchronization session (optional)}
    \sync\xleftarrow{} \{E_{\key}(\initiator,\epoch)|\var{Postamble}\} \phantom{123456} \> \sendmessageright*{\sync} \< \\
        \>                              \< \epoch \xleftarrow{} \fn{syncEpoch}(\sync)
    \pclb \pcintertext[dotted]{End Synchronization session}
    %% ranging
    \pclb \pcintertext[dotted]{Start Ranging session}
    \req \xleftarrow{} H_{\key}(\initiator,\epoch) \> \sendmessageright* {\req}        \<  \\
    \{T^k_{W,n}\} \xleftarrow{} \fn{genDelay}(H_\key,\reflector_k,\epoch,n,W) \> \< \{T^k_{W,n}\} \xleftarrow{} \fn{genDelay}(H_\key,\reflector_k,\epoch,n,W) \\
    \{\resp^k_{n}\} \xleftarrow{} H_\key(\reflector_k,\epoch,n) \> \< \{\resp^k_{n}\} \xleftarrow{} H_\key(\reflector_k,\epoch,n)
        \pclb \pcintertext[center]{(Start Batch Response)}
        \>                                  \< (\text{wait } T^k_{W,0}) \\
        \> \sendmessageleft*{\resp^k_0}     \< \\
        \> \vdots\phantom{12345678912345}   \<  \vdots\phantom{12345}   \\
        \>                                  \< (\text{wait } T^k_{W,n}) \\
        \> \sendmessageleft*{\resp^k_n}     \< 
        \pclb \pcintertext[center]{(End Batch Response)}
    \text{Compute ToF}                                                              \> \< \\
    \batch          \xleftarrow{} \{\resp^k_n\}                                     \> \< \\
    \tilde{\batch}  \xleftarrow{} \fn{scanResponses}(\batch)                        \> \< \\
    ToF_k           \xleftarrow{} \fn{estimateToF}(\tilde{\batch},\{T^k_{W,n}\})    \> \<
    \pclb \pcintertext[dotted]{End Ranging session}
}
\caption{Secure Broadcast Ranging protocol.}
\label{fig:appendix-protocol}
\end{figure*}

\Cref{fig:appendix-protocol} represents our proposed secure broadcast ranging protocol.

\end{document}